\documentclass[usenatbib]{mn2e}

\makeatother
\usepackage[latin1]{inputenc}
\usepackage{multirow}
\usepackage{graphicx}
\usepackage{multicol}
\usepackage{float}
\usepackage{color}
\usepackage{multirow}
\usepackage{footnote}
\usepackage{amssymb}
\usepackage[figuresright]{rotating}
\usepackage{amsmath}
\usepackage{url}
\title[Deep H$\alpha$ observations of NGC~247 and NGC~300]{Deep Fabry-Perot H$\alpha$ observations of two Sculptor group galaxies, NGC~247 and NGC~300}
\author[J. Hlavacek-Larrondo et al.]{J. Hlavacek-Larrondo$^{1,2}$\thanks{E-mail: juliehl@ast.cam.ac.uk}, M. Marcelin$^{3}$, B. Epinat$^{4,5}$, C. Carignan$^{1,6}$, M.-M. de  \newauthor{Denus-Baillargeon$^{1}$, O. Daigle$^{1,3}$, O. Hernandez$^{1}$} \\ 
$^1$Laboratoire d'Astrophysique Exp\'{e}rimentale, D\'{e}partement de physique, Universit\'{e} de Montr\'{e}al, C.P. 6128, Succ. centre-ville, Montr\'{e}al, \\ Qu\'{e}bec, Canada, H3C 3J7\\ 
$^2$Institute of Astronomy, Madingley Road, Cambridge, CB3 0HA\\
$^{3}$Laboratoire d'Astrophysique de Marseille, Universit\'{e} de Provence, CNRS, 38 rue Fr\'{e}d\'{e}ric Joliot-Curie, F-13388 Marseille Cedex 13, \\ France \\ 
$^{4}$Universit\'{e} de Toulouse; UPS-OMP; IRAP; Toulouse, France \\
$^{5}$CNRS; IRAP; 14, Avenue Edouard Belin, F-31400 Toulouse, France \\
$^{6}$Observatoire d'Astrophysique de l'Universit\'{e} de Ouagadougou, BP 7021, Ouagadougou 03, Burkina Faso}

\begin{document}

\date{Accepted 2009 ?. Received 2009 ?; in original form 2009 ?}

\pagerange{\pageref{firstpage}--\pageref{lastpage}} \pubyear{2009}

\maketitle

\label{firstpage}

\begin{abstract}
It has been suggested that diffuse ionized gas can extend all the way to the end of the H\thinspace{\sc i} disc, and even beyond, such as in the case of the warped galaxy NGC~253 \citep{Bla1997490}. Detecting ionized gas at these radii could carry significant implications as to the distribution of dark matter in galaxies. With the aim of detecting this gas, we carried out a deep H$\alpha$ kinematical analysis of two Sculptor group galaxies, NGC~247 and NGC~300. The Fabry-Perot data were taken at the 36-cm Marseille Telescope in La Silla, Chile, offering a large field of view. With almost 20 hours of observations for each galaxy, very faint diffuse emission is detected. Typical emission measures of 0.1 cm$^{-6}$ pc are reached. For NGC~247, emission extending up to a radius comparable with that of the H\thinspace{\sc i} disc ($r\sim13'$) is found, but no emission is seen beyond the H\thinspace{\sc i} disc. For NGC~300, we detect ionized gas on the entirety of our field of view ($r_{\rm max}\sim14'$), and find that the bright H\thinspace{\sc ii} regions are embedded in a diffuse background. Using the deep data, extended optical rotation curves are obtained, as well as mass models. These are the most extended optical rotation curves thus far for these galaxies. We find no evidence suggesting that NGC~247 has a warped disc, and to account for our non detection of H$\alpha$ emission beyond its H\thinspace{\sc i} disc, as opposed to the warped galaxy NGC~253, our results favour the model in which only through a warp, ionization by hot young stars in the central region of a galaxy can let photons escape and ionize the interstellar medium in the outer parts. 
\end{abstract}


\section{\large Introduction}

It has usually been thought that the H\thinspace{\sc i} disc in spiral galaxies extends to radii well beyond the optical disc. H\thinspace{\sc i} rotation curves have therefore led to extensive studies of the dark matter distribution at large radii. 

However, \citet{Bla1997490} suggested that diffuse optical emission could in principle be found at radii larger than the H\thinspace{\sc i} disc. They initially suggested that beyond a certain radius, cold gas could no longer support itself against ionization by the ambient radiation field. To test this theory, they obtained very deep optical spectra for one of the brightest Sculptor group galaxies, NGC~253, and succeeded in detecting ionized gas (H$\alpha$ and [N\thinspace{\sc ii}]) beyond the H\thinspace{\sc i} disc. The rotation curve they derived also showed signs of a possible decline. However, their results led them to conclude that the existence of this extended ionized gas was not due to the ambient radiation field, but more likely due to hot young stars in the central regions of the galaxy ionizing the outer disc through a strong warp present in the galaxy's disc. 

Detecting diffuse ionized gas can therefore not only provide insight as to the kinematics of a galaxy at large radii, but also provide strong constraints on the source of its ionization. In this context, \citet{Dic2008135} carried out a deep H$\alpha$ kinematical analysis of NGC~7793, another Sculptor group galaxy. They used the 36-cm Marseille Telescope located in La Silla, Chile, offering a large field of view. With almost 20 hours of observations, they detected ionized emission all the way to the end of the H\thinspace{\sc i} disc and confirmed that NGC~7793 had a truly declining rotation curve. In order to complete this study, another observing run was carried out in October 2007 with the objective of studying the H$\alpha$ kinematics and extent of the diffuse ionized gas of three other bright Sculptor group galaxies: NGC~253, NGC~247 and NGC~300. The results concerning NGC~253 have been published in \citet[][]{Hla2010}. For this galaxy, we were able to find optical emission reaching out to 1.6 times the maximum radius of the H\thinspace{\sc i} disc. We succeeded not only in detecting ionized gas beyond the detections of \cite{Bla1997490}, but were also able to confirm the declining part of the rotation curve. Our observations also favoured an ionization source by hot young stars in the central regions of a warped galaxy.

We now present the results concerning NGC~247 and NGC~300. For NGC~247, we seek to see if diffuse optical emission can also be seen on a scale similar to the H\thinspace{\sc i} disc, and even beyond. For NGC~300, the H\thinspace{\sc i} disc extends to $r=20'$, but previous H$\alpha$ studies have only been able to detect ionized gas up to $r\sim10'$ \citep{Mar1985151} and down to emission measures (EM) of 6 cm$^{-6}$ pc \citep{Hoo1996112}. Although we will be limited by our field of view for this galaxy, we will be able to reach further distances ($r\sim14'$) and at least an order of magnitude deeper in flux, allowing us to further study the H$\alpha$ kinematics of the galaxy. 

We first present the data (Section 2), and then the velocity and total integrated H$\alpha$ emission maps (Section 3). In Section 4, we derive the kinematical parameters and rotation curves, and we show the mass models in Section 5. The results are discussed in Section 6, and summarized in Section 7. A distance of 1.80 Mpc for NGC~300 and 2.53 Mpc for NGC~247 have been adopted in this study \citep{Puc198895}. The optical parameters of NGC~300 and NGC~247 are summarized in Table \ref{a2_t1a}. 
\begin{table}
\centering
\caption{Optical parameters of NGC~300 and NGC~247 \label{a2_t1a}}
\resizebox{0.64\textwidth}{!}{
\begin{minipage}[t]{14cm}
    \renewcommand{\footnoterule}{}
\begin{tabular}{@{}lll@{}}
\hline
\hline
& NGC~300 & NGC~247 \\
\hline
Morphological type\footnote[1]{\citet{deV1962136} for NGC~300; \citet{Car198558} for NGC~247}  & SA(s)d & SAB(s)d \\
R.A (2000)\footnote[2]{RC3} & 00$^h$54$^m$53.5$^s$ & 00$^h$47$^m$08.5$^s$ \\
Dec. (2000)$^b$ & -37$^o$41$^{'}$04${''}$ & -20$^o$45$^{'}$37${''}$ \\
Isophotal major diameter$^b$, D$_{25}$ (\textit{B}) & 21.9${'}$ & 21.4${'}$  \\
Holmberg radius$^a$, R$_{\rm HO}$ (\textit{B}) & 11.7${'}$ & 12.2${'}$ \\
Exponential scale length$^a$, $\alpha$$^{-1}$ (\textit{B}) & 2.06 kpc & 2.9 kpc \\
Axis ratio$^a$ (q$\equiv$b/a)  & 0.74 & 0.28  \\
Inclination$^a$ (q$_0$ = 0.12), \textit{i} & 42.3$^{o}$ & 75.4$^{o}$ \\
Total apparent \textit{B} magnitude\footnote[3]{\citet{Car198558}}, \textit{B}$^{0,i}$$_{\rm T}$ & 8.70 & 9.67 \\
Total corrected apparent \textit{B} magnitude\footnote[4]{Internal absorption A(\textit{i}) for NGC~300 = 0.21; for NGC~247$=0.68$ (RC3)}\footnote[5]{Galactic extinction A$_{B}$ for NGC~300 = 0.02; for NGC~247$=0.07$ (RC3)}, & 8.47 & 8.92 \\
\indent \textit{B}$^{0,i}$$_{\rm T}$ & & \\
Adopted distance\footnote[6]{\citet{Puc198895}} (Mpc) & 1.80 & 2.53 \\
 & (1$^{'}$ = 0.52 kpc) & (1$^{'}$ = 0.73 kpc)\\
Absolute \textit{B} magnitude, M$^{0,i}$$_{B}$ & -17.81 & -18.09\\
Total blue luminosity (M$_\odot$ = 5.43), L$_B\odot$ & 1.97$\times$10$^{9}$ & 2.57$\times$10$^{9}$ \\
\hline
\end{tabular}
\end{minipage}}
\label{a2_t1a}
\end{table}

\section{\large{Fabry-Perot observations and data reduction}}
The FP observations of NGC~253, NGC~247 and NGC~300 were taken with the 36-cm Marseille Telescope at La Silla, Chile. The reduction steps are the same as those used in the study of NGC~253 in \citet{Hla2010} and are described in this section. 

A commercial camera Andor iXon was used, equipped with an EMCCD (Electron Multiplying Charge Coupled Device) detector \citep[details concerning this detector can be found in][]{Dai20045499,Dai20066276, Dai20087014, Dai2009121}. The Quantum Efficiency (QE) is $\sim90$ per cent at H$\alpha$. The interference filter has a central wavelength of $\lambda_{\rm c}=6565.0$ \AA, and a FWHM of 30 \AA. Maximum transmission was $\sim80$ per cent at 20$^o$C. The FP interference order was ${\rm p}=~765$ at H$\alpha$, with a Free Spectral Range (FSR) of 8.58 \AA\ covering 40 channels of 0.21 \AA\ each. We calibrated the data in wavelength with a neon lamp ($\lambda=6598.95$\AA). The EMCCD has dimensions of $512\times512$ pixels, and an image scale of $2.8''$ pixel$^{-1}$ (total field of view is $23.6'\times23.6'$). The exposure time was 15 seconds per channel. For NGC~300, a total of 115 cycles were observed (1150 minutes or 19.2 hours), whereas for NGC~247 a total of 123 cycles were observed (1230 minutes or 20.5 hours). Since NGC~300 has a larger spatial coverage, it was also necessary to observe separate cubes to extract the sky background. For every 6 cycles taken for NGC~300, $\sim$ 2 cycles of sky were observed (by moving the telescope 20$'$ to the South). 
\begin{figure}
\centering
\begin{minipage}[c]{0.49\linewidth}
  \centering \includegraphics[width=\linewidth]{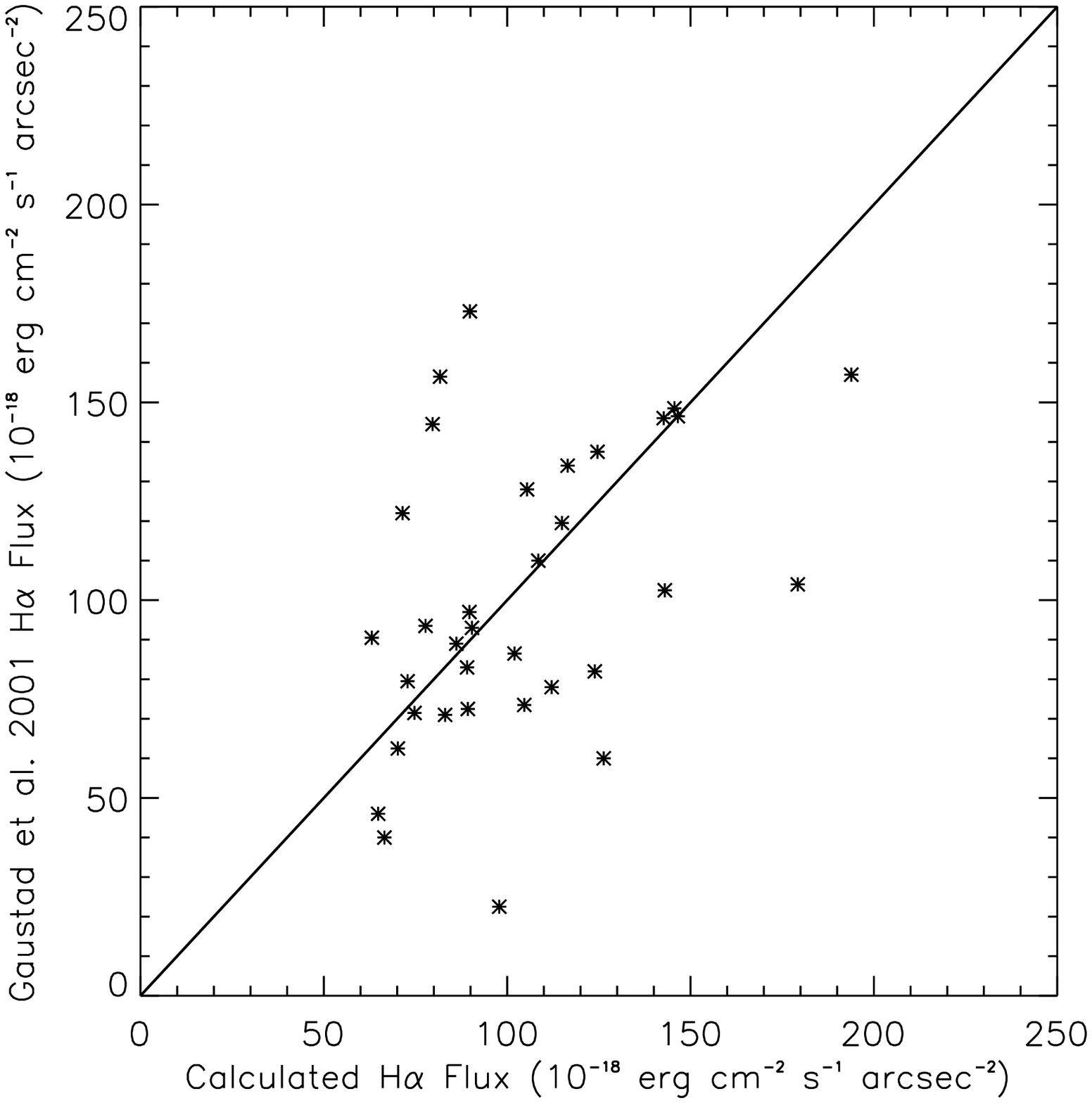}
\end{minipage}
\begin{minipage}[c]{0.49\linewidth}
  \centering \includegraphics[width=\linewidth]{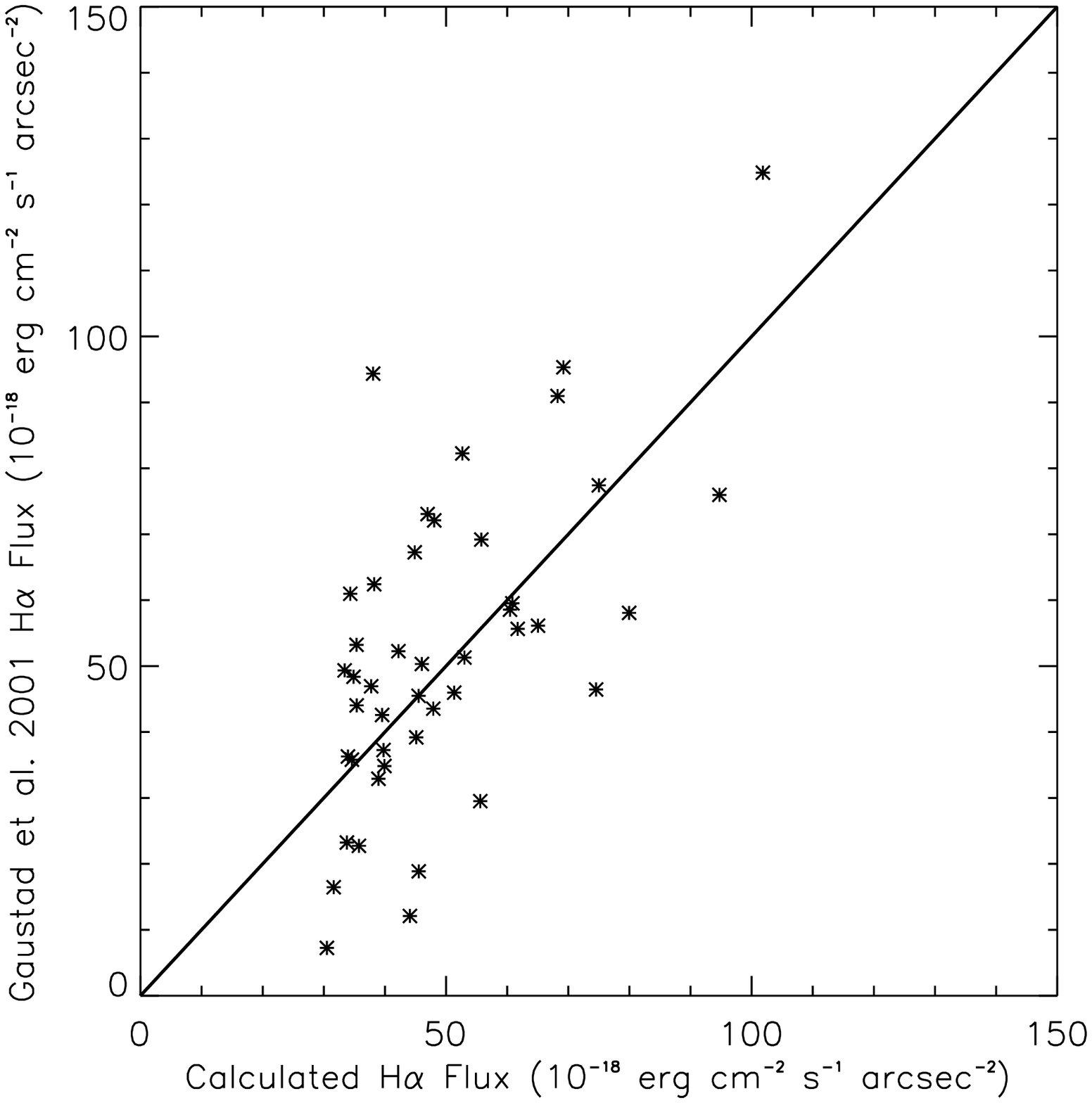}
\end{minipage}
\caption{Comparison between the fluxes obtained for NGC~300 (left) and NGC~247 (right) in this study, and those of \citet{Gau2001113}. Each point represents a bright H\thinspace{\sc ii} region in a galaxy. The continuous line has a slope set to unity.}
\label{a2f1}
\end{figure}

We calibrated in flux the data with the SHASSA (Southern H$\alpha$ Sky Survey) catalogue \citep[][ http://amundsen.swarthmore.edu/]{Gau2001113}, but the catalogue was first corrected for [N\thinspace{\sc ii}] contamination ($\lambda\lambda$ 6548 \AA, 6583 \AA). For NGC~300, the intensity of the [N\thinspace{\sc ii}] contamination was taken as 0.20 times that of H$\alpha$, and for NGC~247, it was taken as 0.24 that of H$\alpha$ \citep{Ken2008178}.  Fig. \ref{a2f1} compares the flux values obtained for NGC~300 (left) and NGC~247 (right) in this study, and those of \citet{Gau2001113}. We estimate that our flux values are within $20-30$ per cent accurate, based on the scatter between our flux values, and those of \citet{Gau2001113}.
 
We now outline the reduction steps. First, we corrected our data for variations of the efficiency on the detector as well as for instrumental transmission variations in the field-of-view using a flat-field observed at twilight. Next, we created wavelength-sorted data cubes corrected for airmass dependence, guiding shifts, cosmic rays, dark, and gain. We then corrected for {\em ghosts}, a common artefact in FP data\footnote[1]{Ghosts are caused by reflections of bright sources (like a bright star or nucleus), and are located symmetrically opposite with respect to the optical axis. They will be at the same wavelength as the primary image, and can thus be well identified, both in position and in velocity.}, using the method of \citet[][]{Epi2008388}. Fourth, we applied a spectral smoothing using a Hanning function, and then sky subtracted the data. We then combined all data cubes over all the cycles and nights observed to create a total integrated cube. A spatial smoothing of $5\times5$ pixels was applied to allow better identification of diffuse gas.  Finally, we fitted the line profiles with a single gaussian component and then corrected for the filter dependence, with a mean temperature of 10$^o$C. We were then able to extract the final velocity, dispersion and H$\alpha$ integrated maps. 
\begin{figure*}
\centering
\begin{minipage}[c]{0.49\linewidth}
  \centering \includegraphics[width=\linewidth]{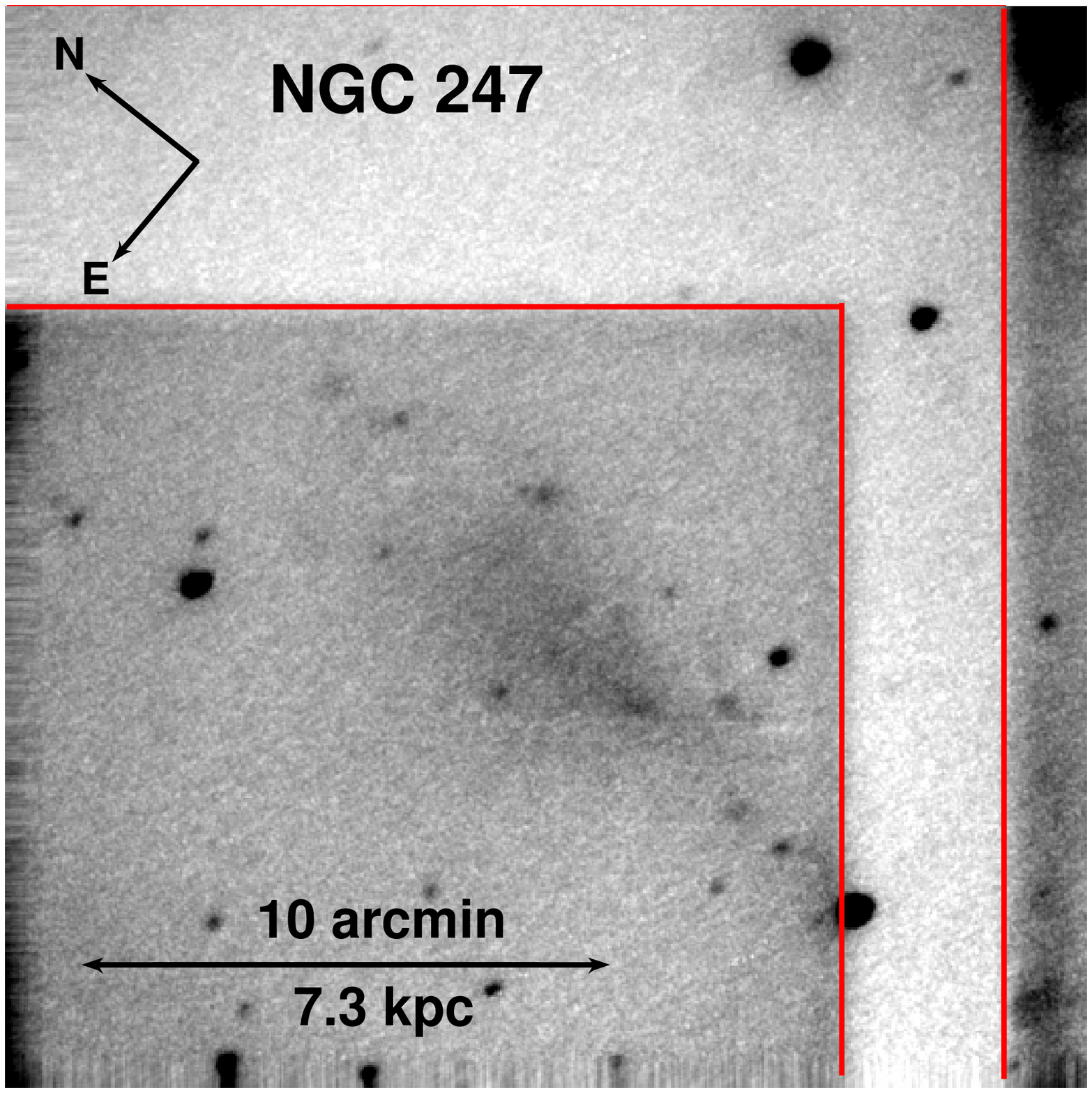}
\end{minipage}
\begin{minipage}[c]{0.49\linewidth}
  \centering \includegraphics[width=\linewidth]{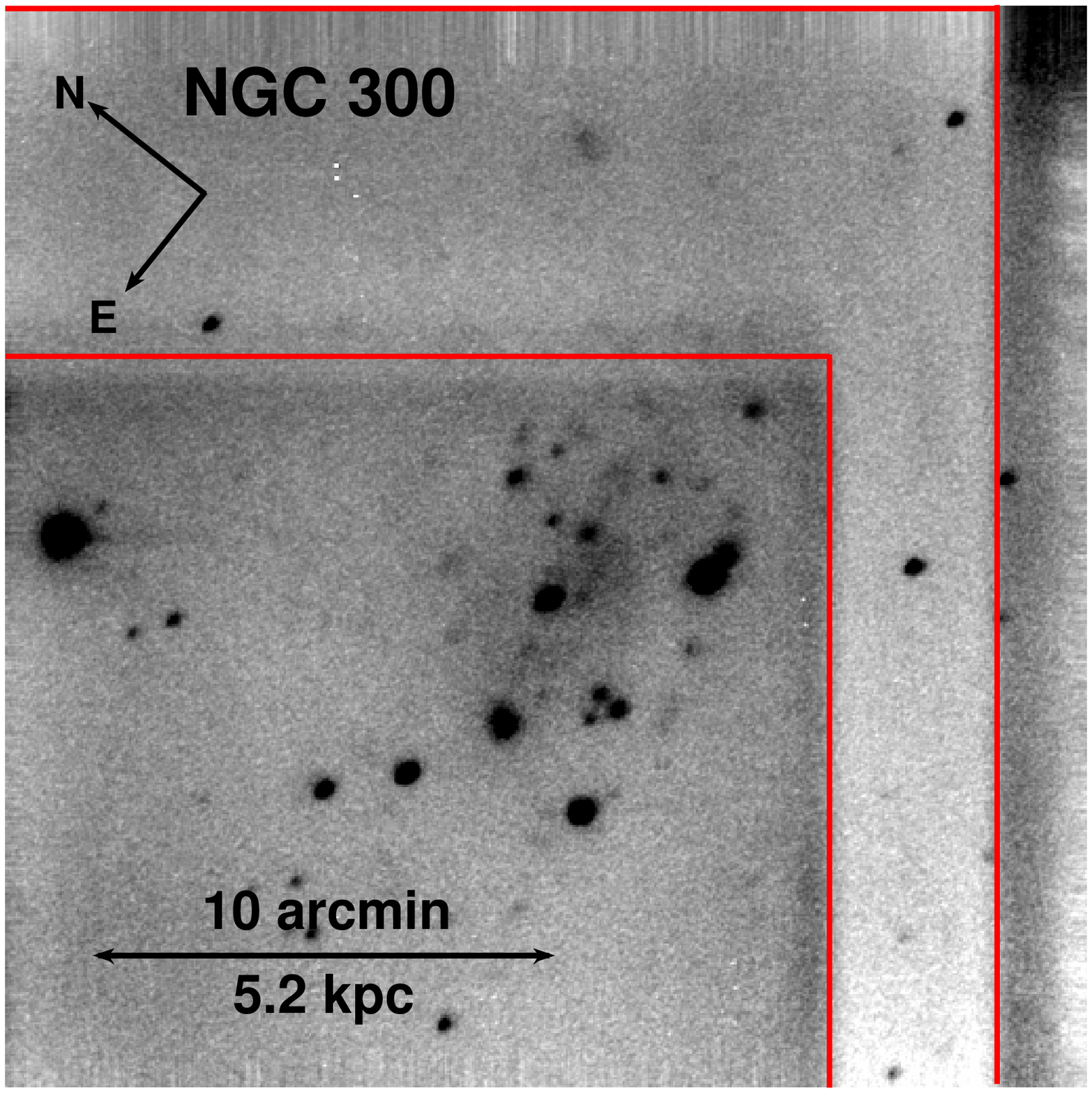}
\end{minipage}
\caption{Field of view of NGC~247 and NGC~300 (linear grey-scale), each totalling $23.6\times23.6'$. These are the raw data sets (summed over 40 channels), before applying the $5\times5$ pixels binning and gaussian fits to the line profiles. Regions contained within red contours show the affected area by the contaminating reflection, and appear less bright because of overcorrection due to the flat-fielding process (see details in the text, at the end of Section 2).The orientation of the north and east axes are shown.}  
\label{a2raw}
\end{figure*}

For each galaxy, we show an image, summed over the 40 channels, of the raw data in Fig. \ref{a2raw} (before applying the $5\times5$ pixel binning and gaussian fits to the spectra). This figure shows that there is a contaminating reflection in the top and right side of the data (see rectangles outlined in red). This reflection arises from light reflected off the detector, and depended on the light present in the dome, and was therefore brighter when taking the flats at twilight than when observing the galaxies. The data cubes were flat-corrected by dividing them with the normalized flat-field image. The continuum of the data cubes, in the regions of reflection, can therefore be inferior to that of the rest of the cube, and is not representative of the true continuum. This parasitic reflection only affected areas of diffuse emission. Section 3 shows in more detail how this reflection was accounted for in the analysis of the diffuse extended emission. 

\section{\large{Velocity fields}}\label{a1c3}

\begin{figure*}
\centering
\begin{minipage}[c]{0.4\linewidth}
  \centering \includegraphics[width=\linewidth]{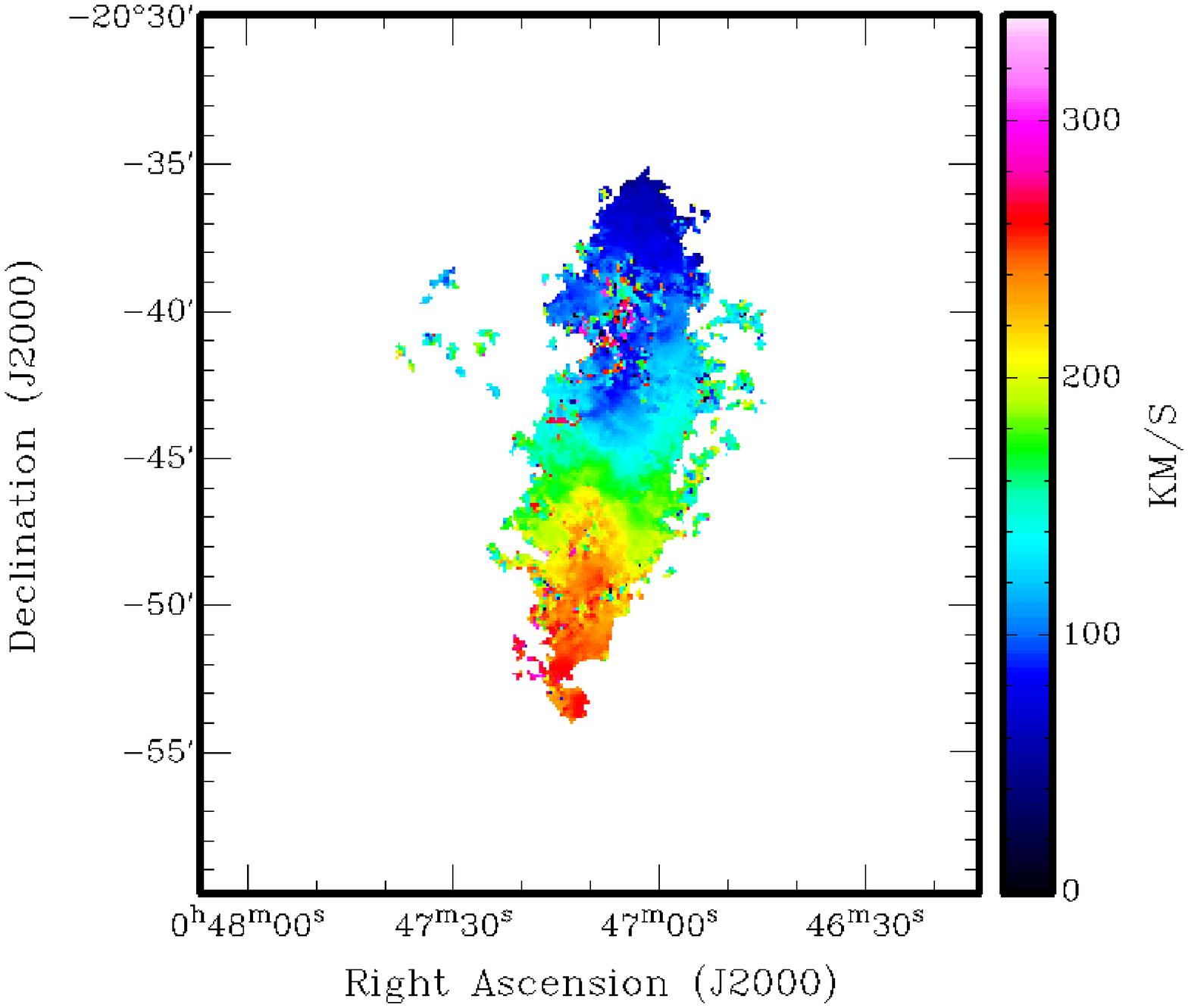}
\end{minipage}
\begin{minipage}[c]{0.4\linewidth}
  \centering \includegraphics[width=\linewidth]{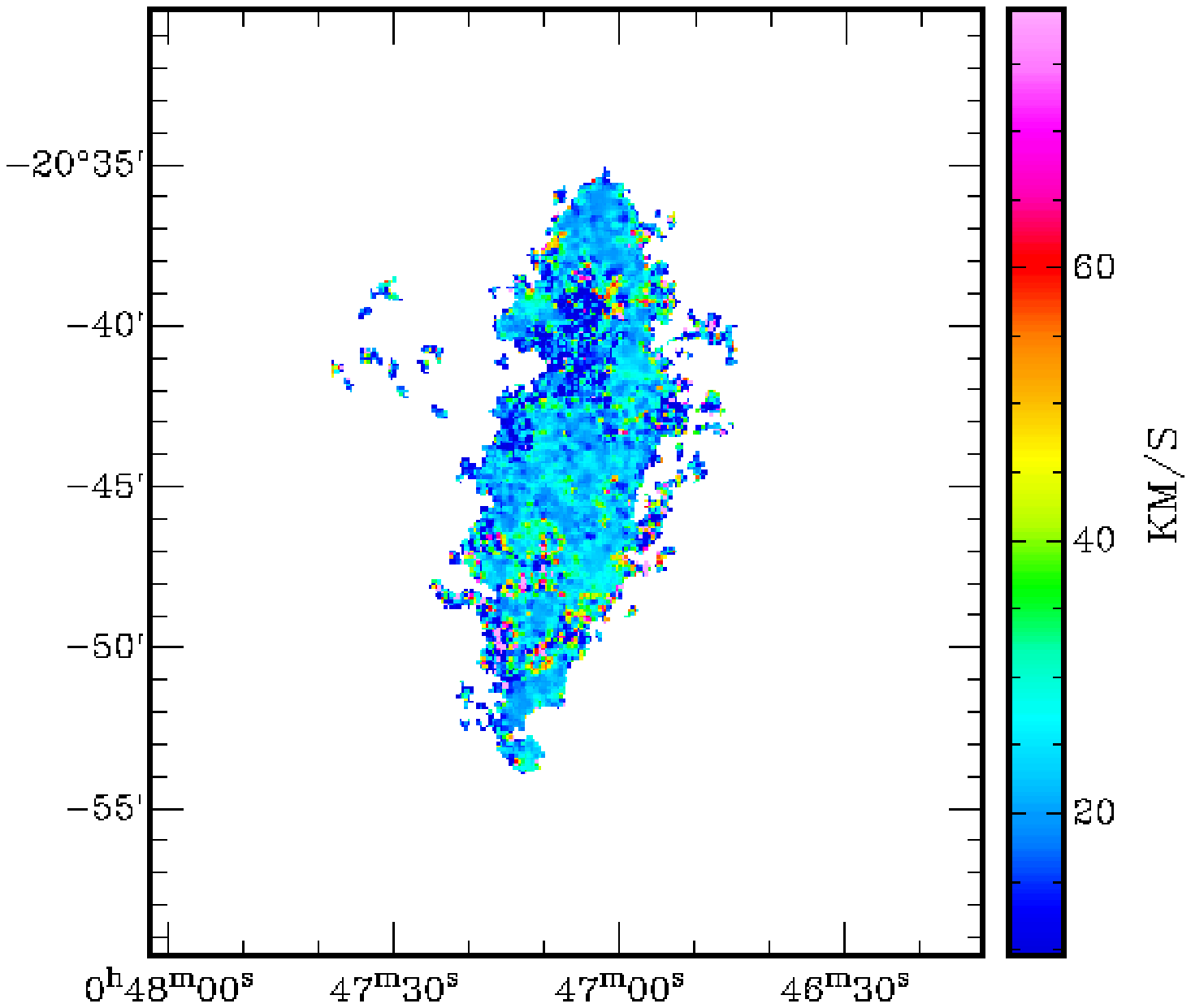}
\end{minipage}
\begin{minipage}[c]{0.4\linewidth}
  \centering \includegraphics[width=\linewidth]{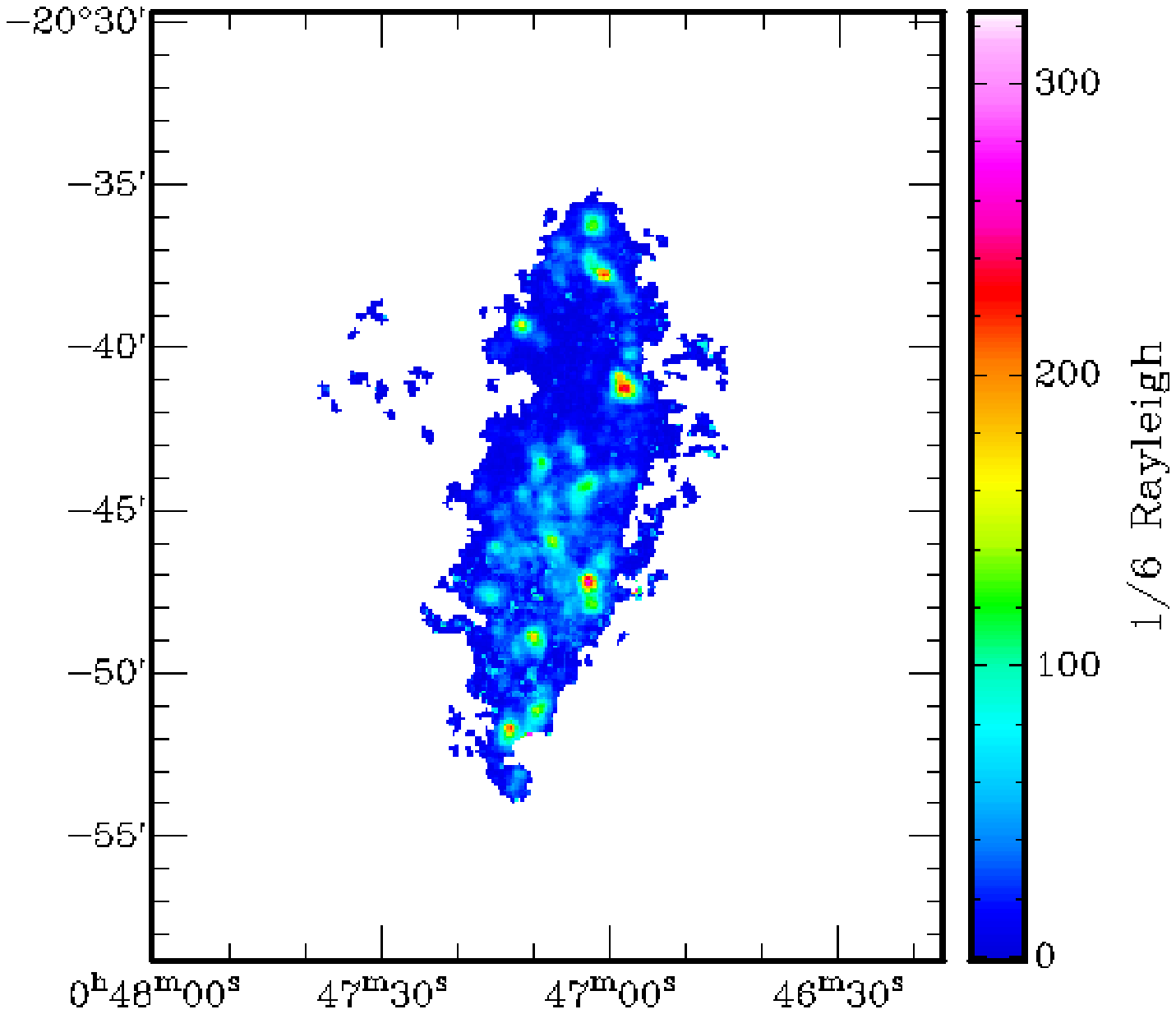}
\end{minipage}
\begin{minipage}[c]{0.4\linewidth}
  \centering \includegraphics[width=\linewidth]{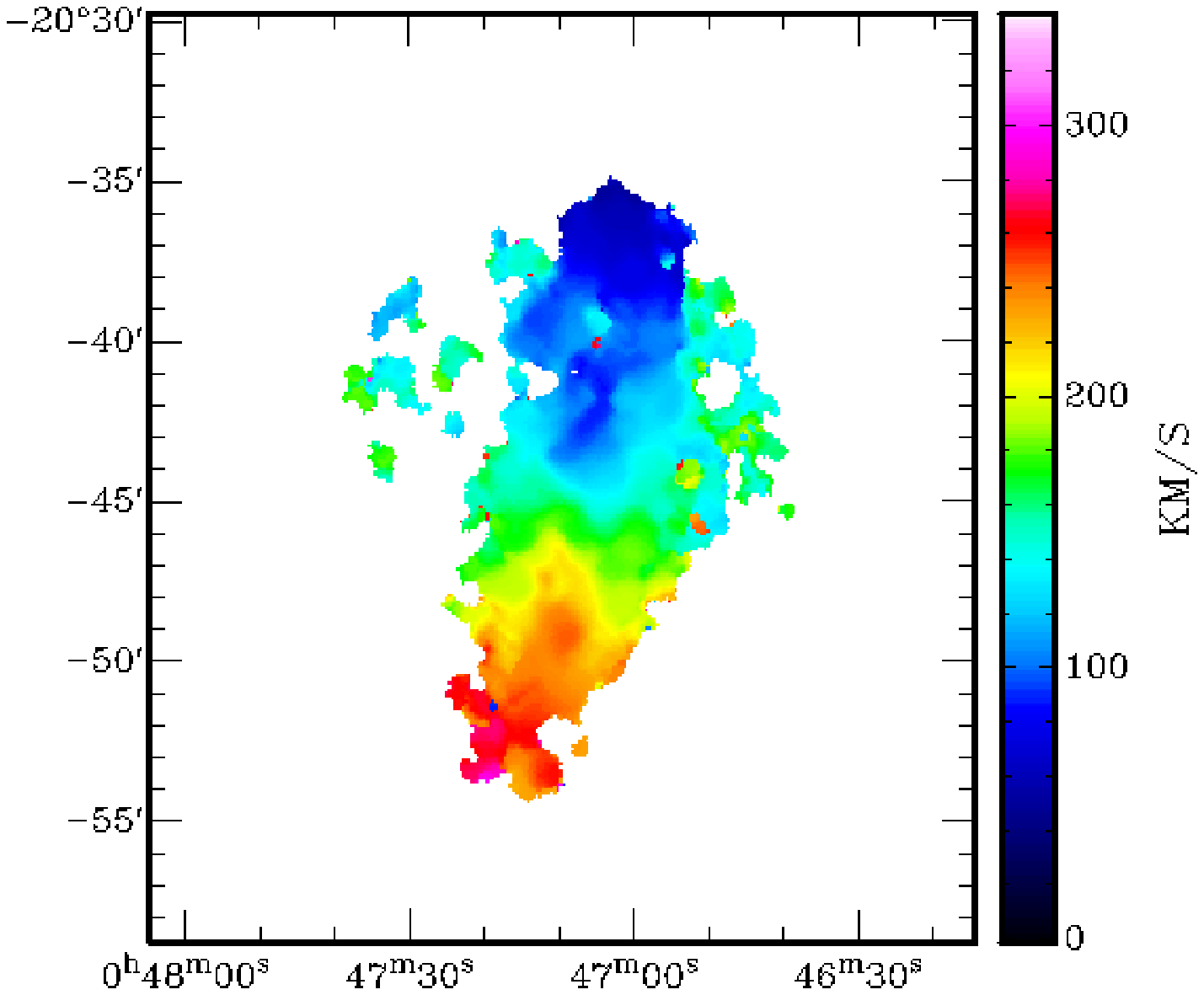}
\end{minipage}
\begin{minipage}[c]{0.5\linewidth}
  \centering \includegraphics[width=\linewidth]{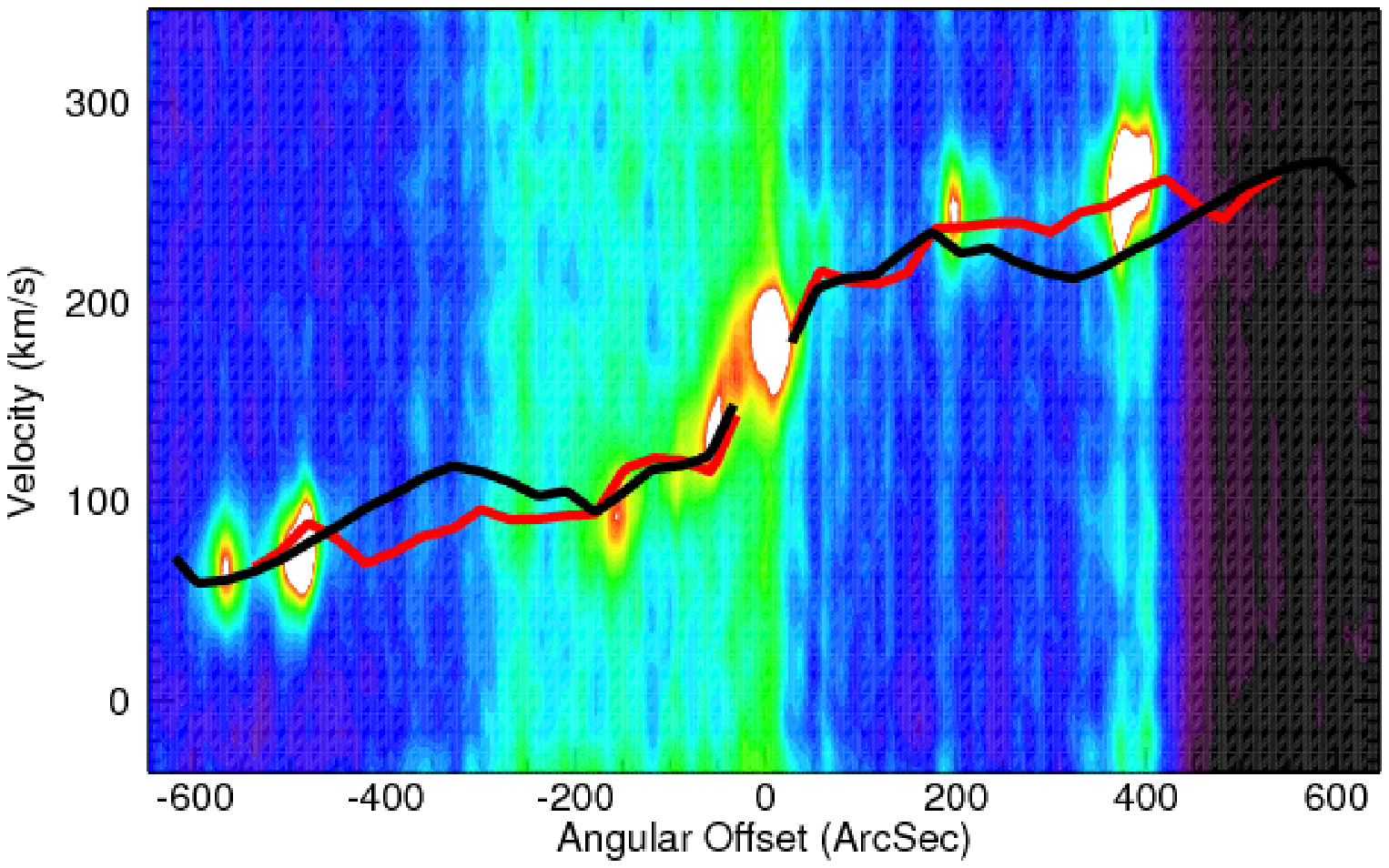}
\end{minipage}
\caption{Deep Fabry-Perot H$\alpha$ kinematical maps of NGC~247 obtained after a $5\times5$ pixels square binning and a single spectral gaussian fitting procedure, unless otherwise specified. Top-left: H$\alpha$ velocity field. Top-right: Dispersion map. Middle-left: Flux calibrated integrated map (scaled in 1/6 Rayleigh or 10$^{-18}$ ergs cm$^{-2}$ s$^{-1}$ arcsec$^{-2}$). Middle-right: H$\alpha$ velocity field obtained with an additional gaussian spatial smoothing of $w_{\rm \lambda}=55''$. Bottom: Position-velocity (PV) diagram along the major axis. The red and black lines are the rotation curves, taken along the major axis (with a slit width of 13 pixels) of a model velocity field. The black line is the rotation curve we derived for both sides, and therefore affected by the non-circular motions of the approaching side. The red line is the rotation curve we derived for the receding side alone, {\rm i. e.} not affected by the non-circular motions. The latter fits much better the maximum intensities of the PV diagram.}
\label{a2f3}
\end{figure*}

\begin{figure*}
\centering
\begin{minipage}[c]{0.45\linewidth}
  \centering \includegraphics[width=\linewidth]{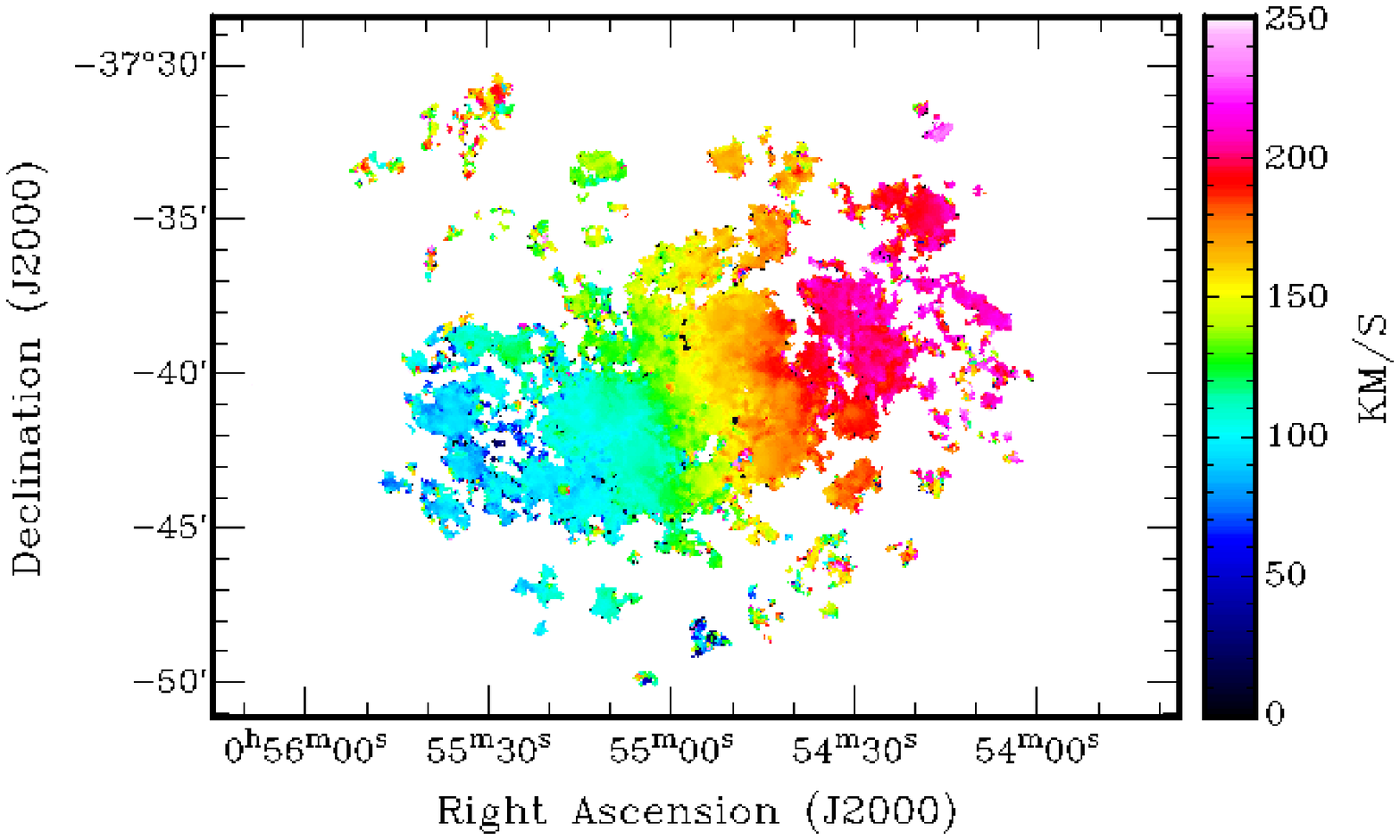}
\end{minipage}
\begin{minipage}[c]{0.45\linewidth}
  \centering \includegraphics[width=\linewidth]{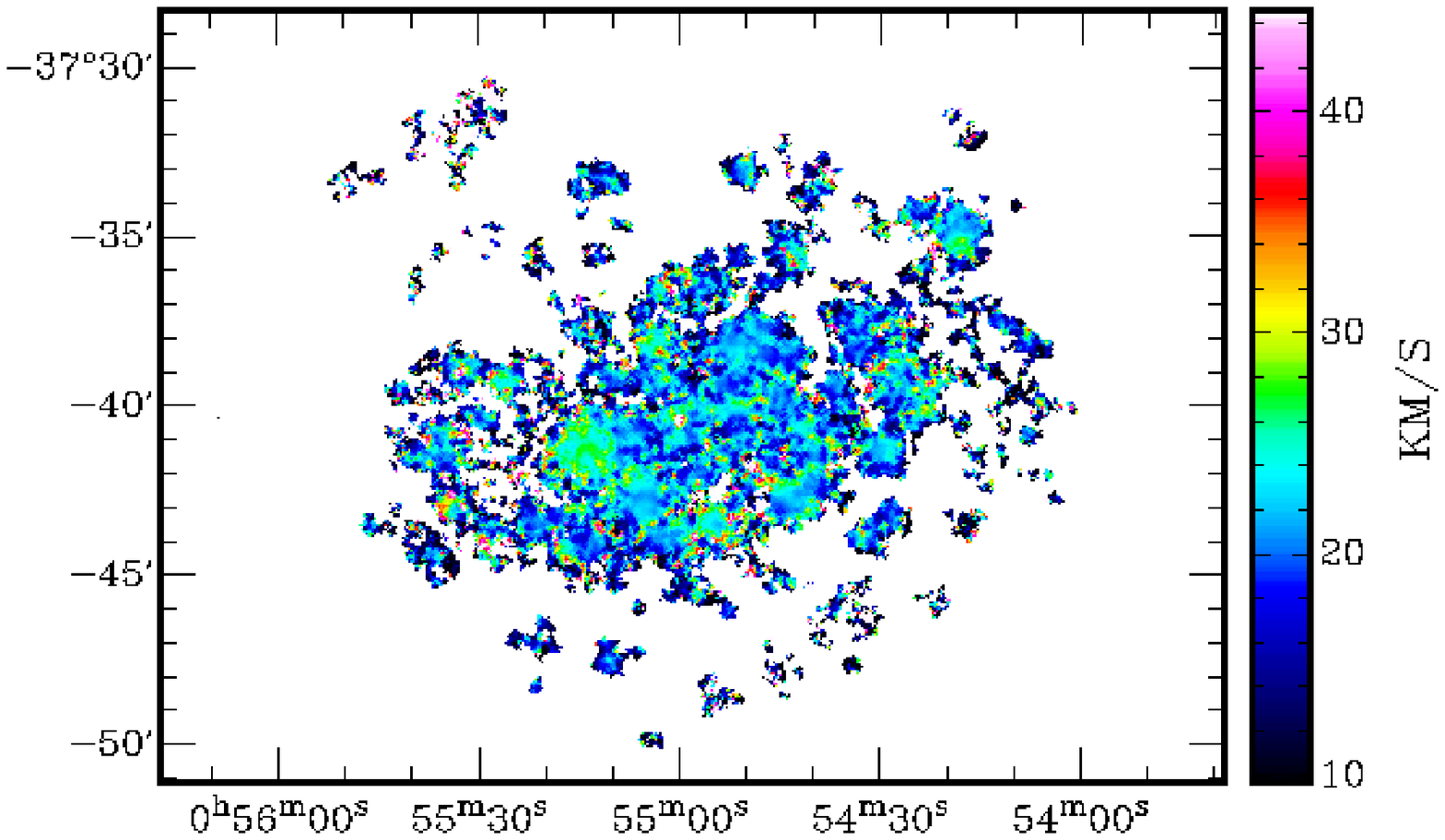}
\end{minipage}
\begin{minipage}[c]{0.43\linewidth}
  \centering \includegraphics[width=\linewidth]{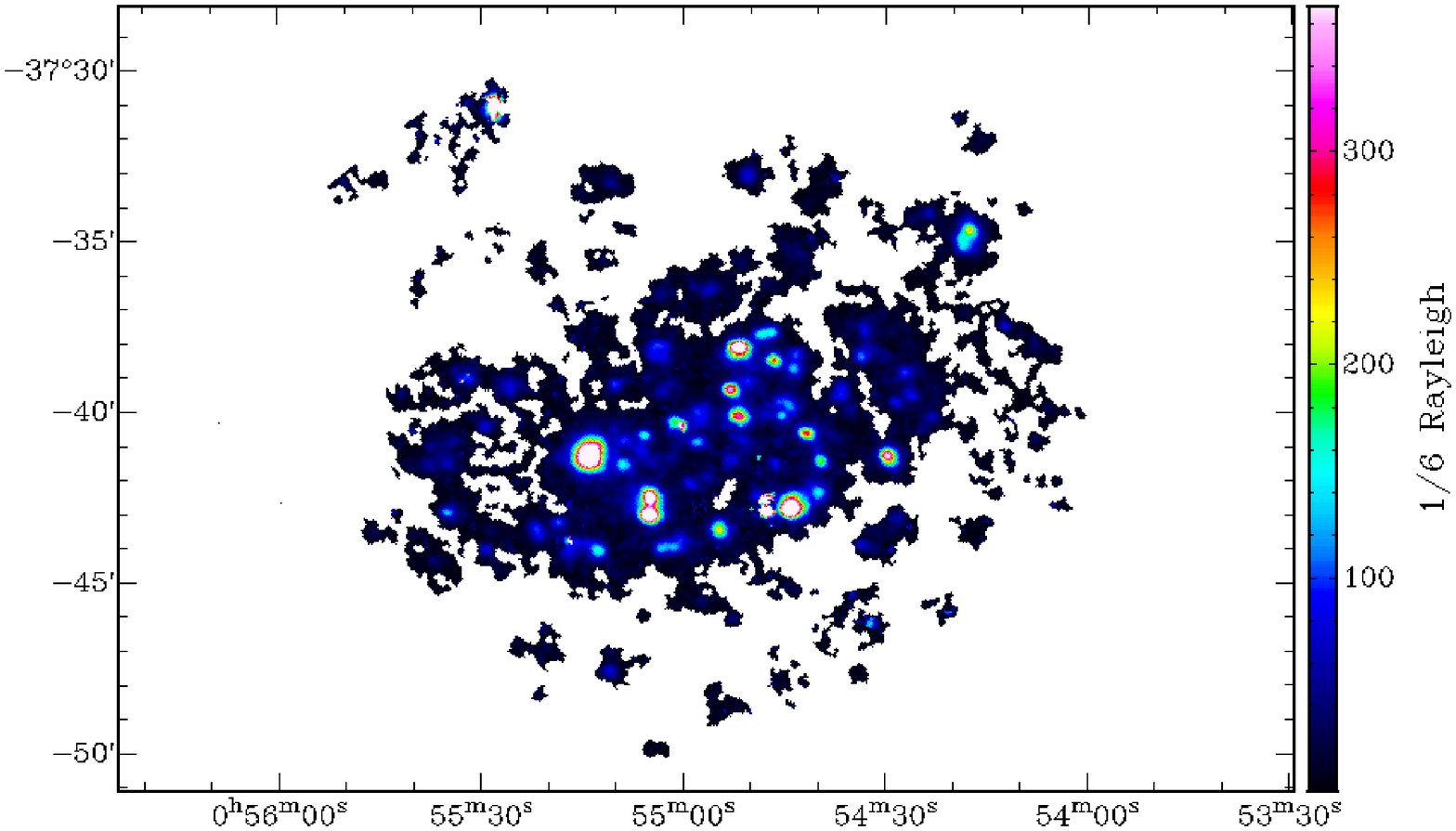}
\end{minipage}
\begin{minipage}[c]{0.45\linewidth}
  \centering \includegraphics[width=\linewidth]{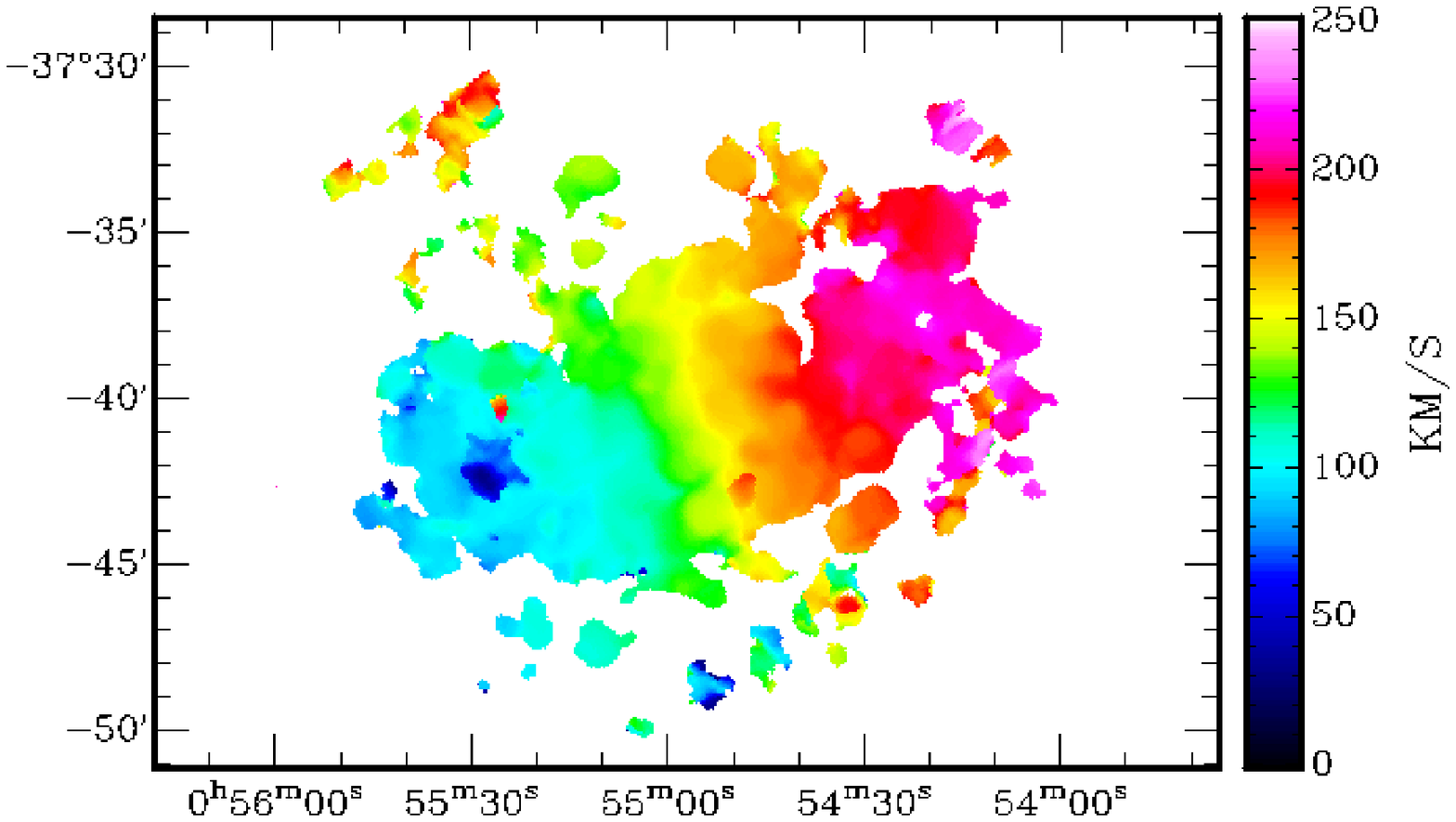}
\end{minipage}
\begin{minipage}[c]{0.5\linewidth}
  \centering \includegraphics[width=\linewidth]{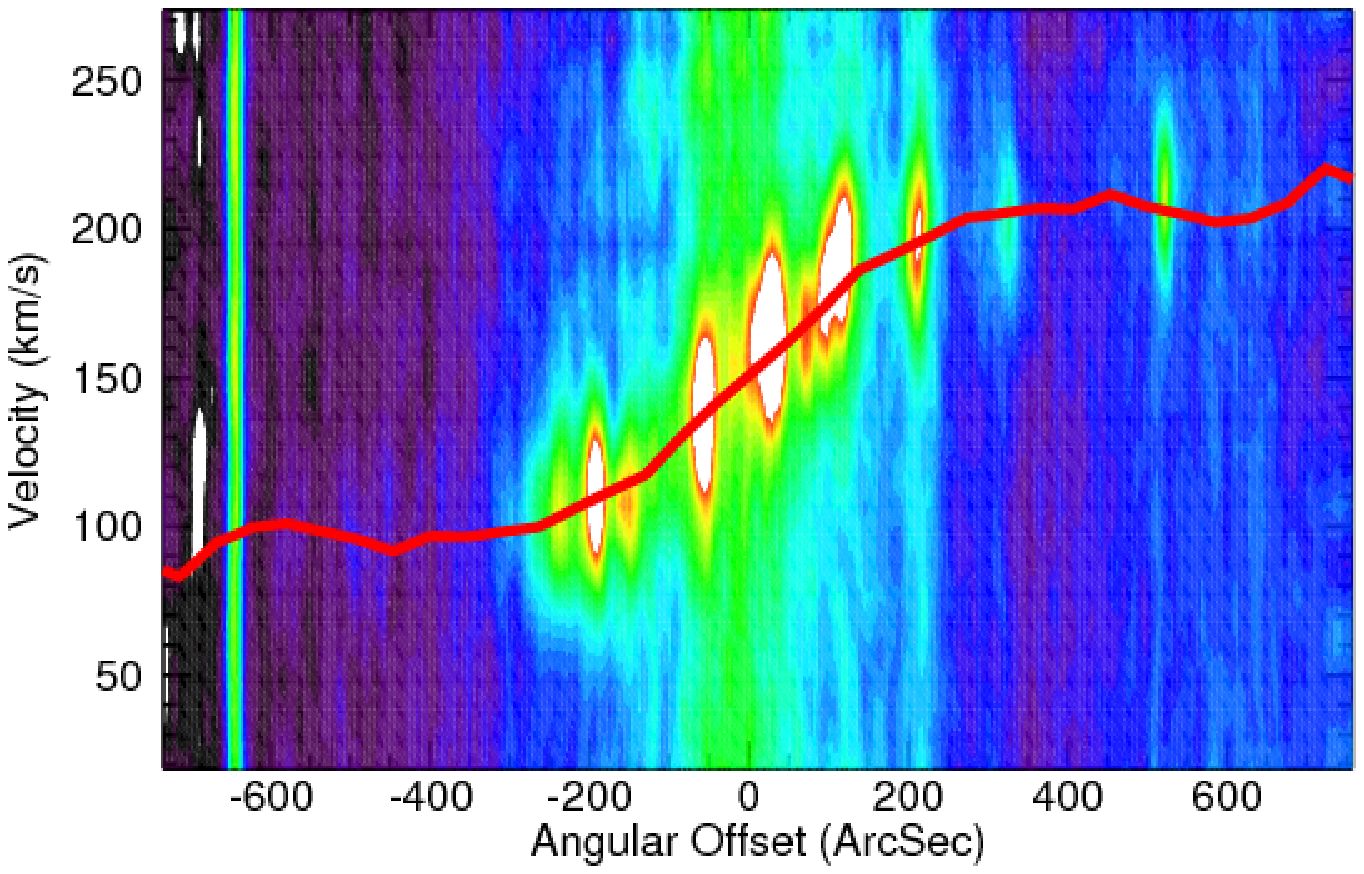}
\end{minipage}
\caption{H$\alpha$ kinematical maps obtained for NGC~300, same caption as Fig. \ref{a2f3}. Top-left: H$\alpha$ velocity field. Top-right: Dispersion map. Bottom-left: Flux calibrated integrated map (scaled in 1/6 Rayleigh or 10$^{-18}$ ergs cm$^{-2}$ s$^{-1}$ arcsec$^{-2}$). Bottom-right: H$\alpha$ velocity field with a gaussian spatial smoothing of $w_{\rm \lambda}=55''$. Bottom: PV diagram along the major axis. The rotation curve shown in red is the one derived for both sides, and is the curve we adopted for the mass models. }
\label{a2f4}
\end{figure*}

We applied the reduction steps outlined in Section 2, and for each emission line we obtained the radial velocity measurement, standard deviation and integrated flux by fitting a gaussian function. These respectively gave the full resolution velocity fields, dispersion maps and flux-calibrated H$\alpha$ integrated maps shown in Fig. \ref{a2f3} (NGC~247) and Fig. \ref{a2f4} (NGC~300). We also obtained lower resolution velocity fields (see Fig. \ref{a2f3} and Fig. \ref{a2f4}, middle-right panels), by applying an additional gaussian spatial smoothing of $w_{\rm \lambda}=55''$, where $w_{\rm \lambda}$ is the width of the smoothing function. These low resolution maps were used to extract the kinematical parameters of the galaxies, which would have been difficult to do with the original full resolution velocity fields because of perturbations in the field ({\rm e. g. see velocity field of interarm region in NGC~247, where we were able to detect some gas}). By using these less disturbed kinematical parameters, we then used the original full resolution velocity fields to extract the rotation curves.

Excessive smoothing could induce a bias in the determination of the inclination towards lower values. \citet{Bos1978} determined that a rotation curve starts to properly trace the kinematics of a galaxy when the scale of the galaxy is more than 7 times the resolution of the data. Since our smoothing corresponds to less than 1/10 of the scale of the galaxies, the values we determine for the inclination should not be affected significantly. 


\subsection{Annular ring binning for NGC~247}

\begin{figure*}
\centering
\begin{minipage}[c]{0.35\linewidth}
  \centering \includegraphics[width=\linewidth]{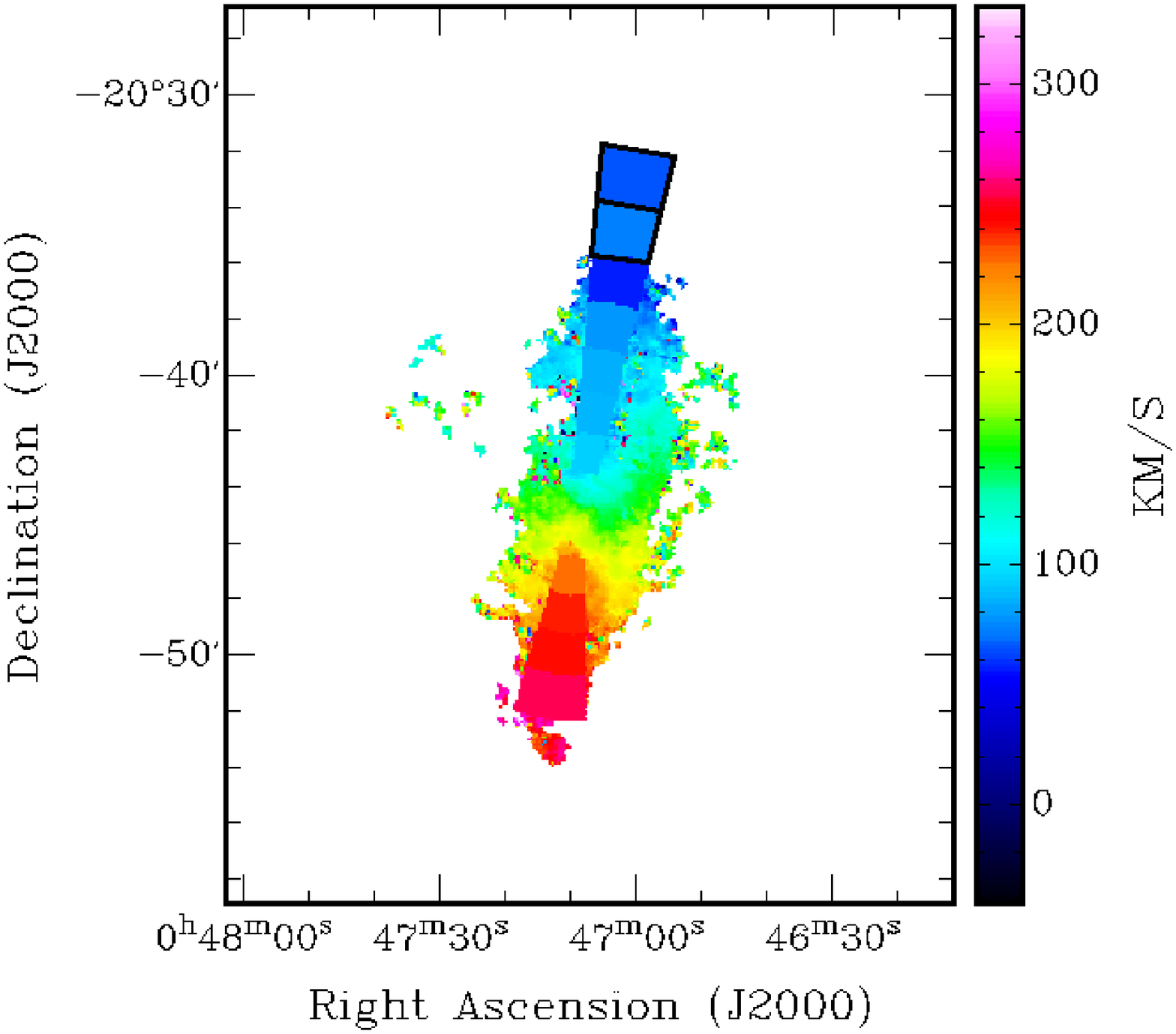}
\end{minipage}
\begin{minipage}[c]{0.32\linewidth}
  \centering \includegraphics[width=\linewidth]{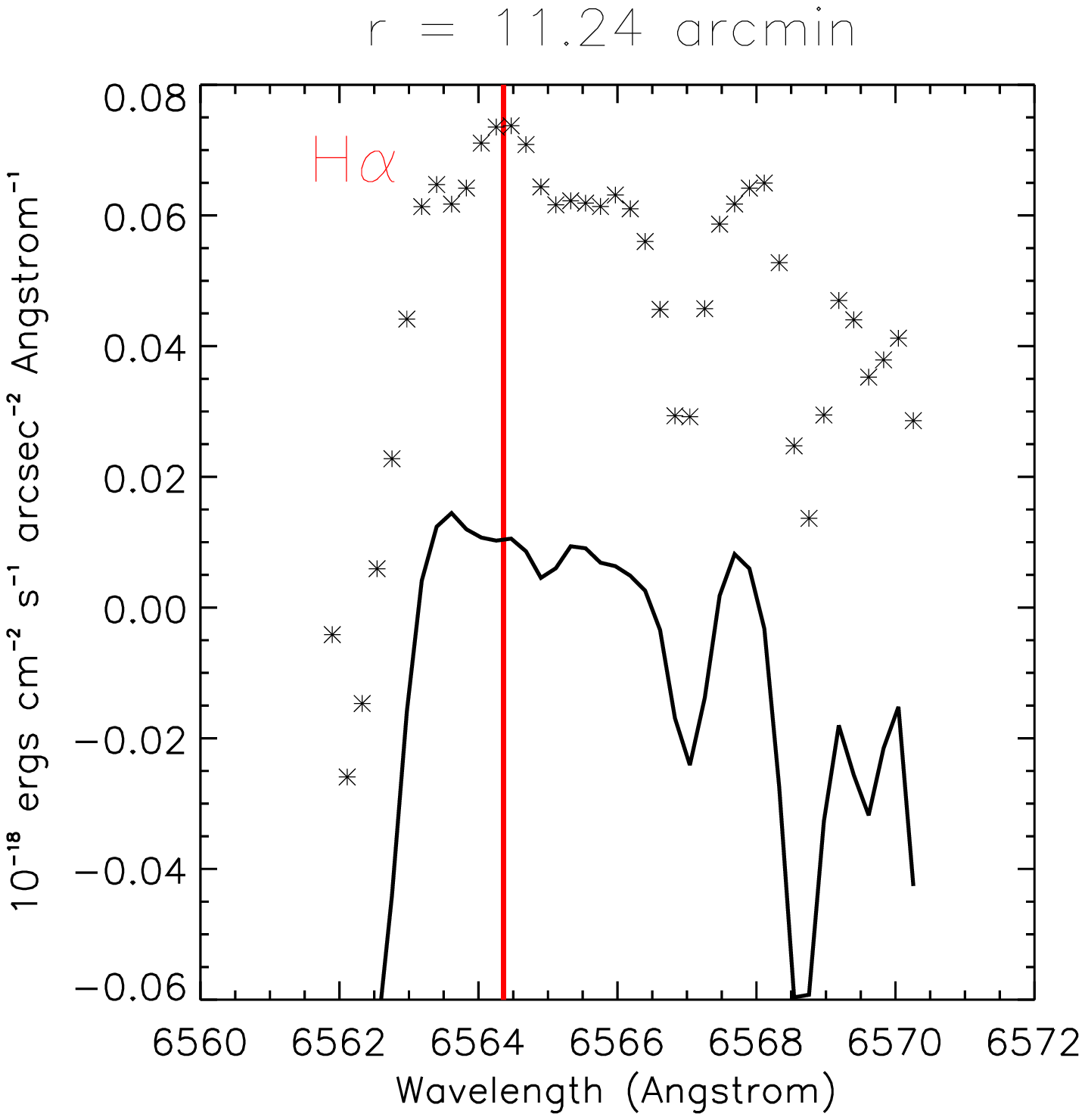}
\end{minipage}
\begin{minipage}[c]{0.32\linewidth}
  \centering \includegraphics[width=\linewidth]{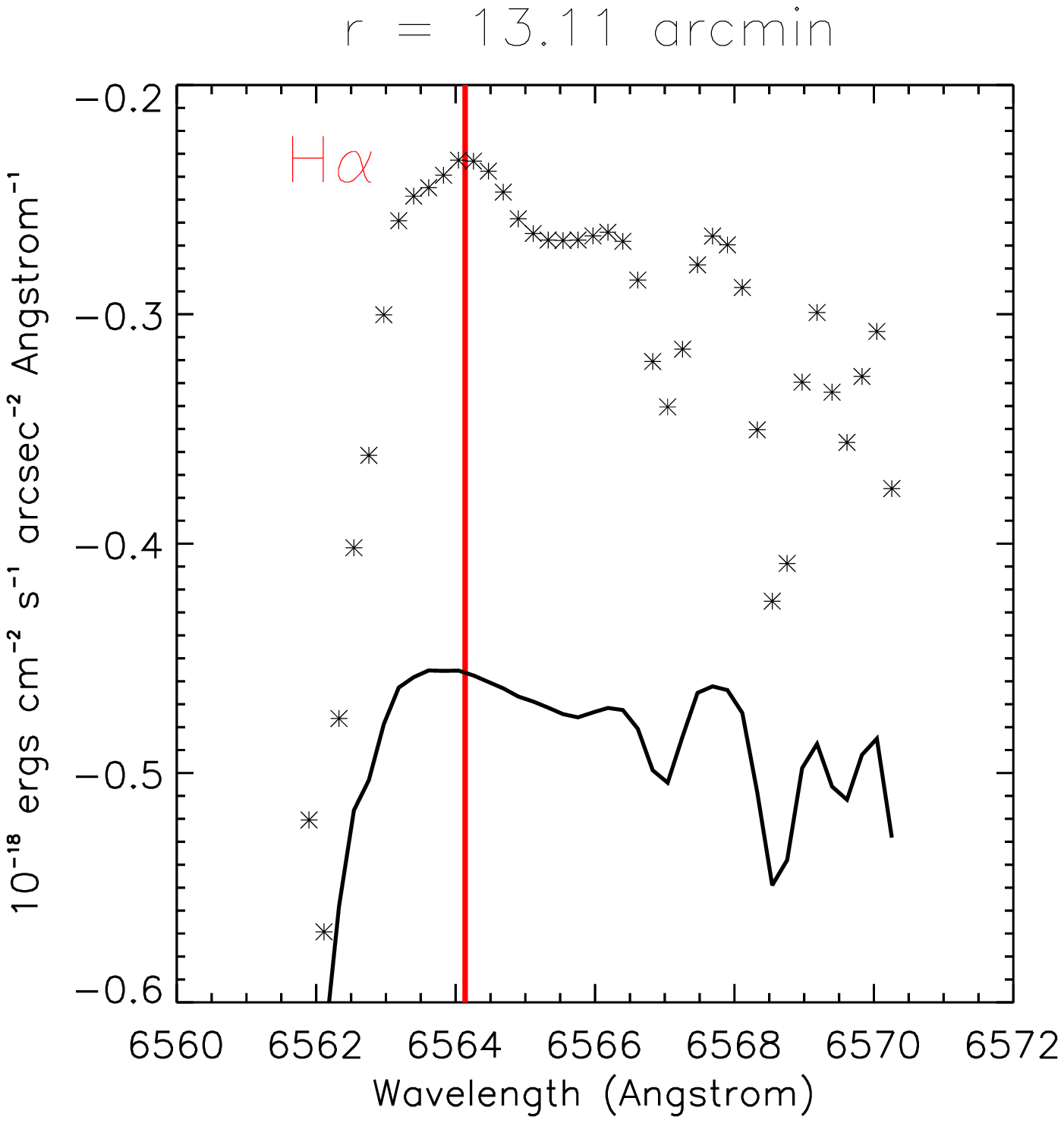}
\end{minipage}
\caption{Left: Velocity field of NGC~247 obtained after a 5 pixels $\times$ 5 pixels square binning procedure, annular ring binning, and a single gaussian adjustment to the emission profiles. The last two binned intervals (black contours) of the northern side show very faint and diffuse H$\alpha$ emission. Middle and right: Spectra of the last two binned intervals of the annular rings in order of radius. The star-like points are for the spectra of the diffuse emission (opening angle of $\sim$ 20$^o$) and the continuous lines are for the contaminating reflection spectra (opening angle of $\sim$ 80$^o$). The red vertical lines represent the location of the H$\alpha$ emission. For the graph centred at a radius of 13.1$'$, the parasitic reflection spectra (continuous line) was lowered in order to allow better comparison with the diffuse spectra (star-like points) by a value of 0.5 in units of 10$^{-18}$ ergs cm$^{-2}$ s$^{-1}$ arcsec$^{-2}$  \AA$^{-1}$. As mentioned in Section 2 and Section 3, the continuum of the spectra is not representative of their true value.}
\label{a2f5}
\end{figure*}

Both \citet{Bla1997490} and \citet{Hla2010} succeeded in detecting diffuse ionized gas in NGC~253 by binning the data into annular rings. Since NGC~247 is also highly inclined ($i\sim74^o$), we can apply the same kind of binning. The annular rings were centred along the major axis in the plane of the image and about the galaxy's dynamical centre. For a given opening angle, different regions corresponding to different radial intervals were binned together. We chose an opening angle of $\sim20^o$ ({\rm i. e.} $10^o$ on each side of the major axis), so that we would not overlap on too large a portion of sky, and made the binning intervals grow with radius, since diffuse emission becomes more difficult to detect as the radius increases (see Fig. \ref{a2f5}). 

As seen from our raw data in Fig. \ref{a2raw}, the extended part of the northern arm of NGC~247 (beyond a radius of 11$'$) fell directly into the areas affected by the contaminating reflections. Here, it became crucial to determine if any emission line detected was the result of diffuse emission or of the reflection. The continuum observed in the affected regions was also not representative of its true value, and could amount to a negative value since it had been previously overcorrected for during sky subtraction. However, the spectrum of the reflection did not vary significantly. By applying another annular ring binning, this time with opening angle of $80^o$ to smooth sufficiently over the affected areas, we were able to obtain a typical response spectrum for the contaminating reflection. If an emission line was only seen in the first (opening angle of 20$^o$) and not the second (opening angle of 80$^o$) binning, then it was considered to be truly associated with diffuse emission. Using this method, we were able to identify two extended radial intervals on the northern side of NGC~247 which had faint H$\alpha$ emission ($r=11.2'$, $r=13.1'$).

\section{\large{Kinematical parameters and rotation curve}}
Four kinematical parameters must be determined from the velocity field to derive the rotation curve. They are the dynamical centre ($x_{\rm 0}$ and $y_{\rm 0}$), the systemic velocity $V_{\rm sys}$, the inclination $i$, and the position angle PA of the major axis. We use the \textsc{gipsy} program and \textsc{rotcur} task to derive them. These programs are based on a least square fitting method that best reproduces the observed velocity field \citep{Beg1989223}. 

Using the lower resolution velocity fields ($w_{\rm \lambda}=55''$), we first find the dynamical centre and $V_{\rm sys}$ by keeping $i$ and PA fixed. The initial values are chosen as those derived in the H\thinspace{\sc i} study (Carignan \& Puche 1990 for NGC~247; Puche et al. 1990 for NGC~300)\nocite{Car1990100,Puc1990100}. To minimize errors due to deprojection effects, we exclude data in an opening angle of 60$^o$ around the minor axis for NGC~247 and 35$^o$ for NGC~300. We use a cosine-square weighting function on the remaining data of NGC~247, where each point is weighted by $\cos(\theta)^2$ with $\theta$ being the angle from the major axis. This maximizes the weight of the points around the major axis. For NGC~300, since the inclination is smaller, we use a simple $\cos(\theta)$ weighting function. 

We find $V_{\rm sys}=164.8\pm5.3$ km s$^{-1}$ for NGC~247 and of $V_{\rm sys}=151.6\pm4.2$ km s$^{-1}$ for NGC~300, in agreement with the systemic velocities found in the H\thinspace{\sc i} studies ($\sim$161 km s$^{-1}$ for NGC~247, and 144.5$\pm$3.0 km s$^{-1}$ for NGC~300). The dynamical centre we derive for NGC~247 coincides with the photometric one, but the one for NGC~300 was slightly shifted by $\sim$ 30$''$ ($\sim$ 260 pc) towards the north-east. In the World Coordinate System, we find that the dynamical centre is located at $\alpha$=00:54:50.4 $\delta$=-37:39:48.0 for NGC~300, and at $\alpha$=00:47:08.6 $\delta$=-20:45:27.3 for NGC~247.

\begin{figure*}
\centering
\begin{minipage}[c]{0.47\linewidth}
  \centering \includegraphics[width=\linewidth]{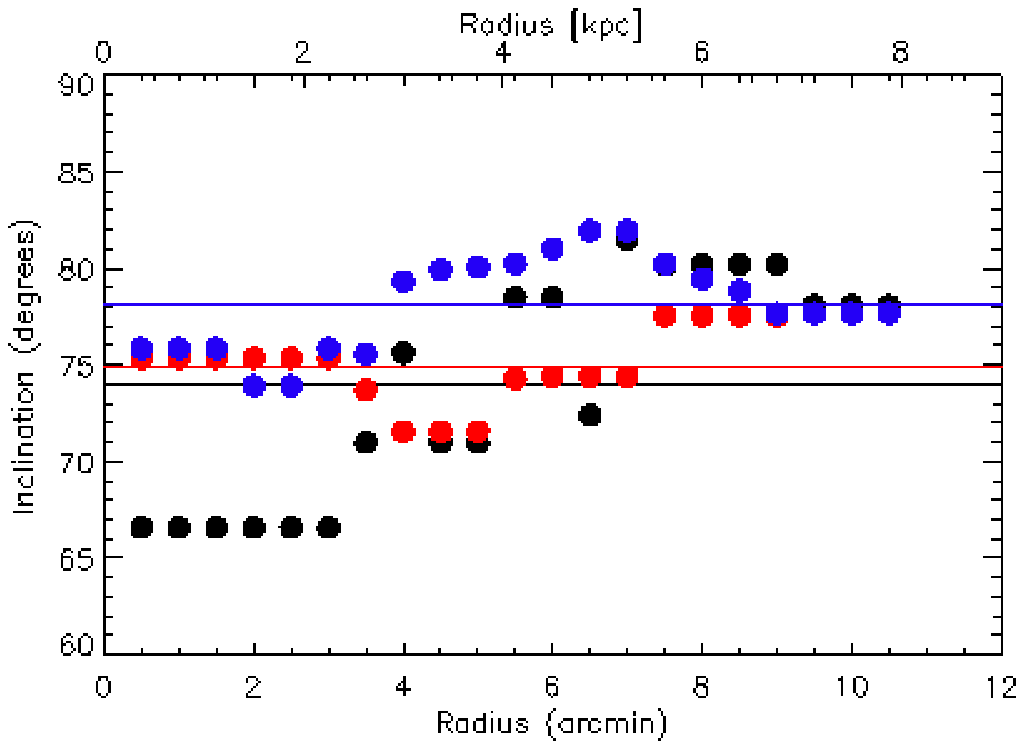}
\end{minipage}
\begin{minipage}[c]{0.47\linewidth}
  \centering \includegraphics[width=\linewidth]{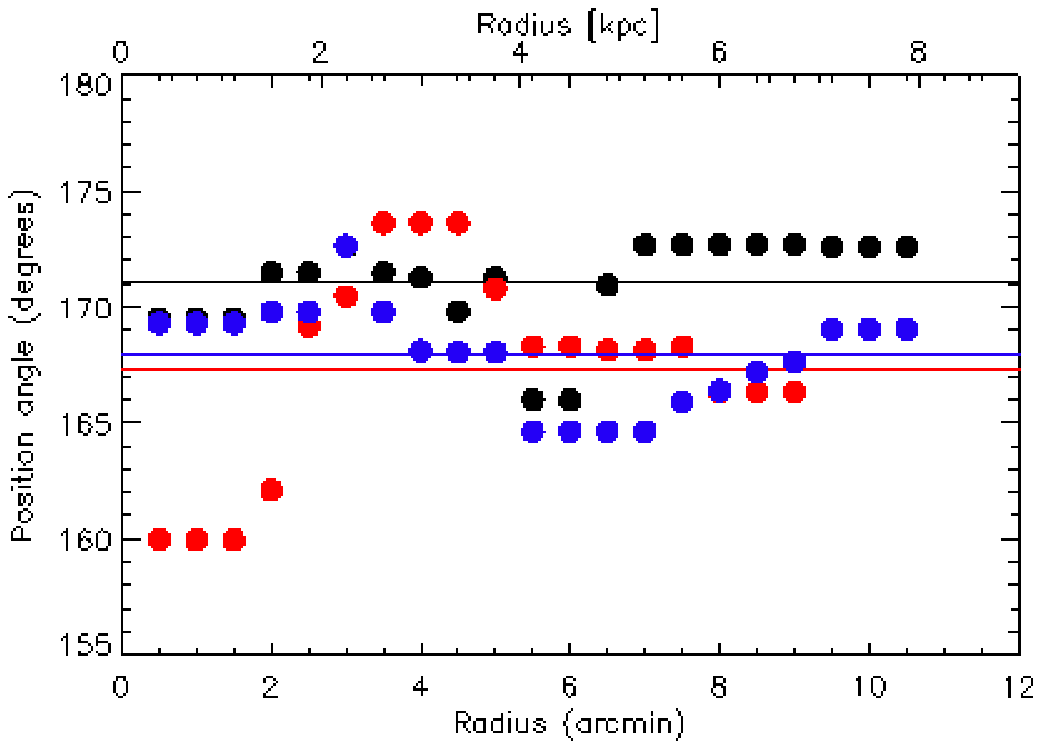}
\end{minipage}
\begin{minipage}[c]{0.47\linewidth}
  \centering \includegraphics[width=\linewidth]{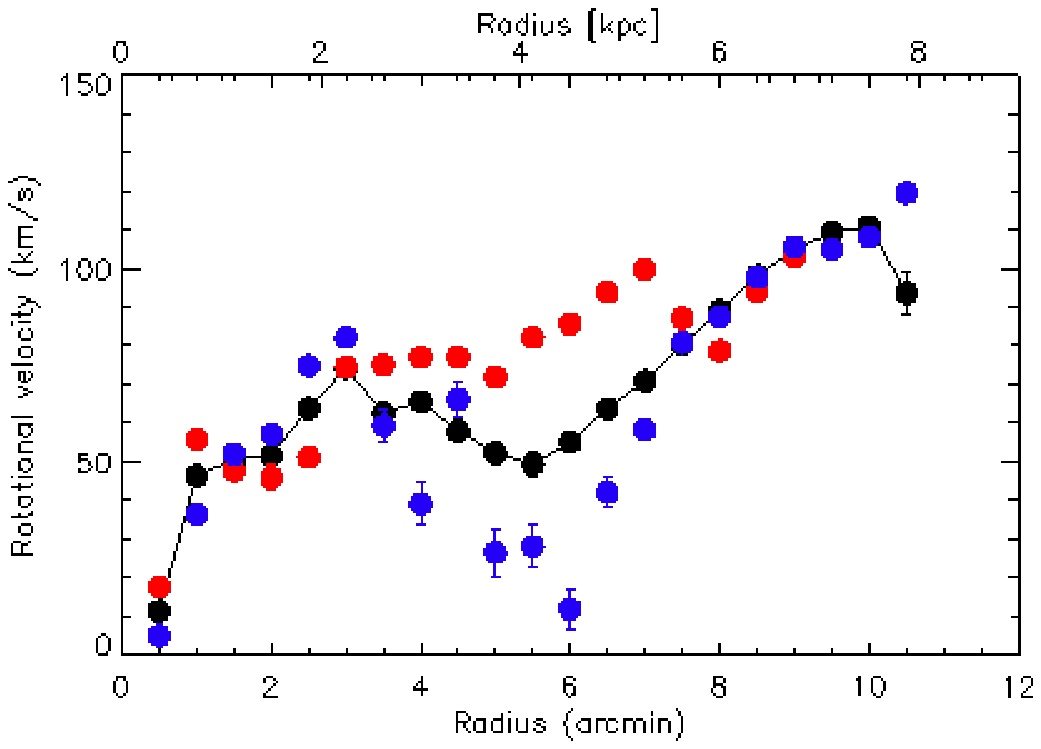}
\end{minipage}
\begin{minipage}[c]{0.47\linewidth}
  \centering \includegraphics[width=\linewidth]{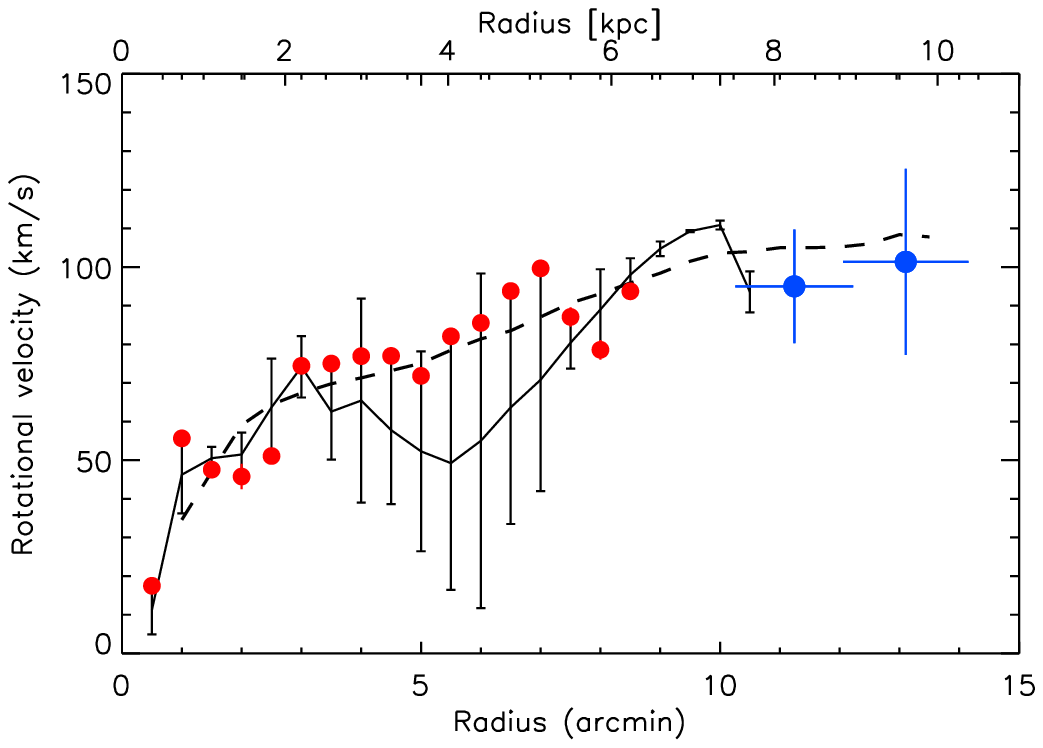}
\end{minipage}
\caption{Top: Median filtered kinematical parameters of NGC~247 derived from the lower resolution velocity field for both sides (black symbols), {\rm i. e.} solutions found considering data points along both receding and approaching sides, as well as separately for the receding (red symbols) and for the approaching (blue symbols) sides. The average values are shown with the horizontal lines. Bottom-left: Rotation curve derived using the average kinematical parameters of the top figures, with a systemic velocity of $V_{\rm sys}=164.8$ km s$^{-1}$, and the full resolution velocity field. The continuous line, along with the derived error bars, represents the solutions found for both sides, while the receding and approaching sides are represented by the red and blue symbols, respectively. Bottom-right: Rotation curve derived for both sides, with the proper error bars (continuous line). These error bars reflect the large difference arising between the velocities derived for both sides, and those of the approaching side, the latter being strongly affected by non-circular motions. They therefore do not represent a large uncertainty in the intrinsic value derived for the velocity. The dashed line represents the H\thinspace{\sc i} data from \citet{Car1990100}. The coloured points illustrate the final adopted rotation curve, and include the two detections of diffuse H$\alpha$ emission in blue.}
\label{a2f6}
\end{figure*}

\begin{figure*}
\centering
\begin{minipage}[c]{0.47\linewidth}
  \centering \includegraphics[width=\linewidth]{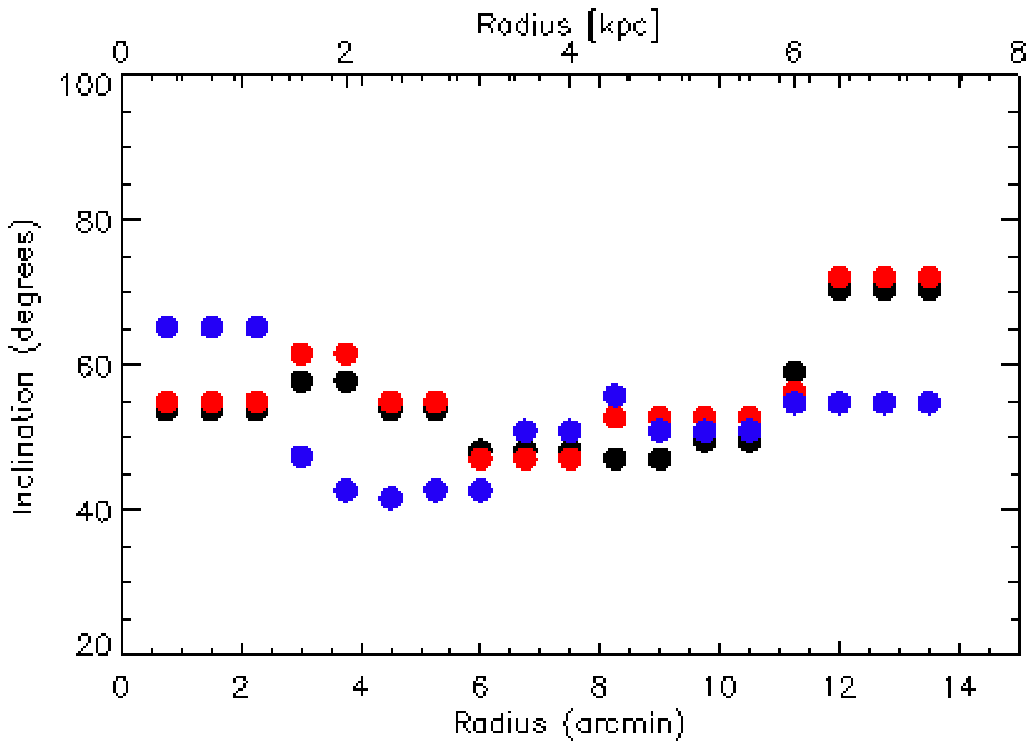}
\end{minipage}
\begin{minipage}[c]{0.47\linewidth}
  \centering \includegraphics[width=\linewidth]{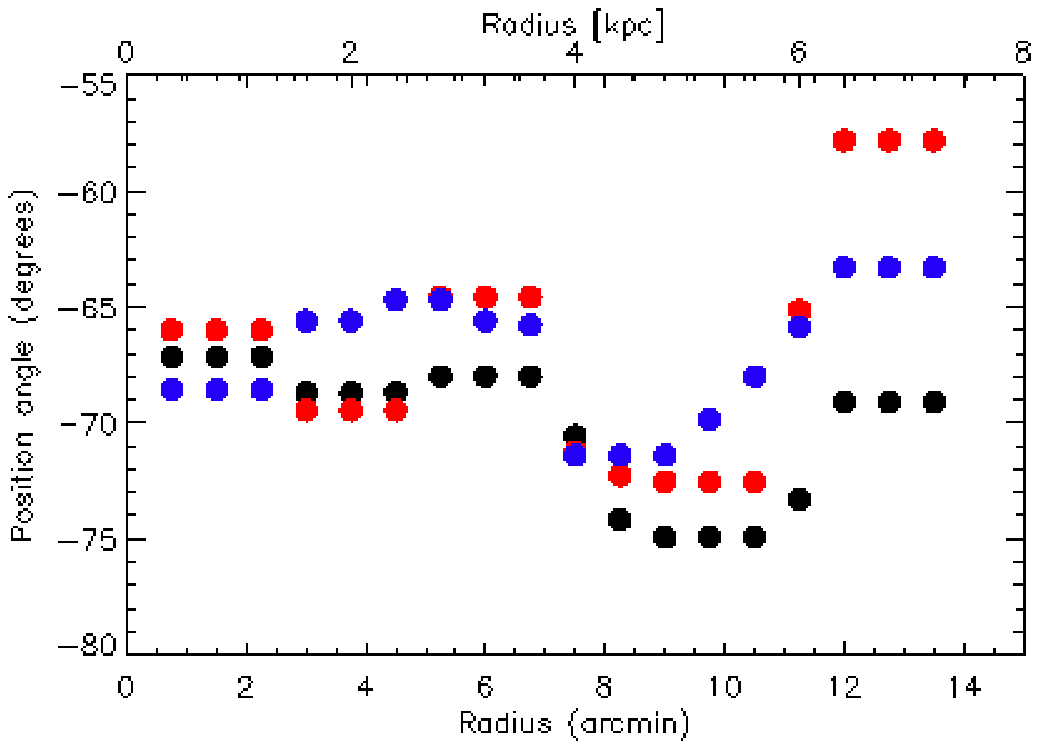}
\end{minipage}
\begin{minipage}[c]{0.47\linewidth}
  \centering \includegraphics[width=\linewidth]{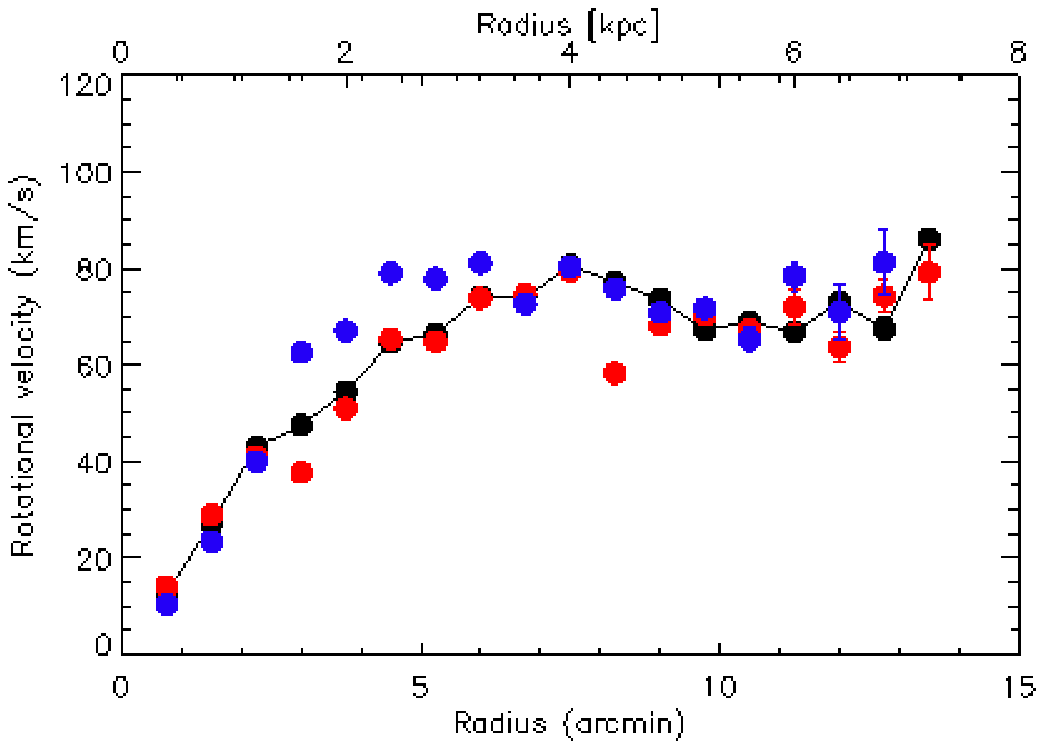}
\end{minipage}
\begin{minipage}[c]{0.47\linewidth}
  \centering \includegraphics[width=\linewidth]{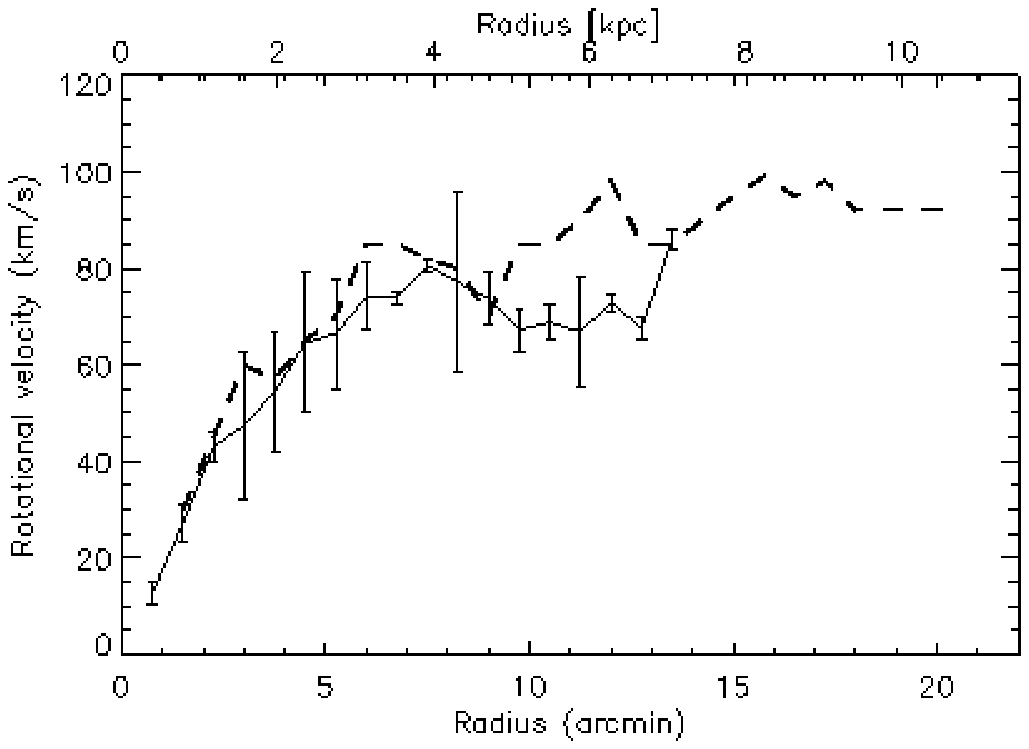}
\end{minipage}
\caption{Same as the previous figure (Fig. \ref{a2f6}), but for the galaxy NGC~300. The systemic velocity of $V _{\rm sys}=151.6$ km s$^{-1}$ was used, and we used the median filtered values of $i$ and PA to derive the rotation curve, therefore allowing some variation with radius to account for the warp. The dashed line represents the H\thinspace{\sc i} data from \citet{Puc1990100}.}
\label{a2f7}
\end{figure*}

Next, we determine $i$ and PA, while the dynamical centre and $V_{\rm sys}$ are kept fixed at the extracted values. Separate solutions are obtained for the receding and approaching side ({\rm i. e.} using data points only on the receding or approaching side of the galaxy). For NGC~247, we find on average that $i=74.0\pm5.8^o$ and PA$=171.0\pm2.0^o$, and for NGC~300, $i=55.1\pm8.0^o$ and PA$=-70.1\pm2.9^o$. {\sc rotcur} could not converge for a small number of radii, and the value of the parameters at these radii were interpolated using adjacent values. For NGC~300, the velocity field extends up to 13.5$'$, but we were only able to extract parameters up to 12.0$'$ (even in the low resolution map, it was difficult to extract reliable kinematical parameters beyond $12.0'$). To derive the complete rotation curve up to a radius of 13.5$'$, we extrapolated $i$ and PA as the ones of the last data point beyond 12.0$'$. Some large fluctuations were seen in the derived values of $i$ and PA, and to reduce their impact when deriving the rotation curve, we applied a median filter to the parameters (median between 5 points). Fig. \ref{a2f6} and Fig. \ref{a2f7} shows the resulting values. Separate solutions are found for the receding, approaching and combined sides. For NGC~300, the variations of $i$ and PA at $r>9'$ follow the same trend as in the H\thinspace{\sc i} data and are consistent with the warp mentioned in \citet[][]{Car1990100}. For NGC~247, more gas is seen on the northern (or approaching) side, and beyond a radius of 8.5$'$, only emission on the approaching side was found. 

Finally, using the full resolution velocity fields, we derive the rotation curves. Since NGC~247 does not show any sign of strong perturbations such as a warp or a bar, we derive the rotation curve using the average value of $i$ and PA, {\rm i. e.} we use the same value of $i$ and PA throughout all radii, but derive the average value from the receding, approaching and combined sides separately. The average values of $i$ and PA are shown in Fig. \ref{a2f6} with the coloured lines. To account for the warp in NGC~300, we must allow some variation of $i$ and PA with radius, and therefore choose to derive the rotation curve using the median filtered values of $i$ and PA, for each side separately. The bottom-left panels of Fig. \ref{a2f6} and Fig. \ref{a2f7} show the solutions found for the approaching and receding side, and the bottom-right panels show the rotation curve derived for both sides, along with the proper error bars. The error bars are defined as the largest difference between the rotation curve of the whole galaxy and that of one of the sides, or if larger, the intrinsic error determined by the \textsc{rotcur} task. The rotation velocities derived for both sides are given in Table \ref{a2_t2} for NGC~247, and in Table \ref{a2_t4} for NGC~300. 

NGC~300 shows no signs of strong non-circular motions. Thus, we simply use the rotation curve derived for both sides as the final adopted curve. However, NGC~247 shows strong perturbations in the rotation curve between $r\sim4'$ and $r\sim8'$, on the approaching side. We associate these with non-circular motions of the interarm region on the approaching side of the galaxy (see Section 6.2.3 for a more detailed look at this region). Applying a mass model to this distorted rotation curve proves to be very difficult, and does not give parameters representative of the true gravitational potential. Instead, we use the rotation curve derived for the receding side as the final adopted rotation curve of the galaxy, and include the two detections of faint H$\alpha$ emission seen in the extended radial intervals at $r=11.2'$ and $r=13.1'$. Table \ref{a2_t3} and the coloured points in the bottom-right panel of Fig. \ref{a2f6} show the values of the final adopted rotation curve. 

The velocities for the emission seen in the two extended radial intervals were derived using the kinematical parameters at maximum radius ($r=10.5'$). For these points, along the x-axis, the error bar was chosen as the half width of the radial interval from which the emission line was extracted. Along the y-axis, we can not estimate the error bars in the same way as the other points ({\rm  i. e.} as the largest difference between the rotation curve of the whole galaxy and that of one of the sides or, if larger, the intrinsic error) since we only detect emission on the approaching side of the galaxy. The error bars were therefore chosen as the dispersion of the gaussian profile fitted to the emission line, since the largest uncertainty is on the location and shape of the profile, rather than on the kinematical parameters. An error on the order of 20$^o$ for the inclination would be required to affect significantly the derived rotation velocities. Finally, the error bars associated with the velocities determined for the receding side (red points in the bottom-right panel of Fig. \ref{a2f6}), were taken as those derived by the \textsc{rotcur} task.

In the bottom panels of Fig. \ref{a2f3} and Fig. \ref{a2f4}, we show the position-velocity (PV) diagrams for each galaxy, along with the final adopted rotation curves in red, and for NGC~247, we include in black the rotation curve derived for both sides, which is affected by the non-circular motions.

\begin{table}
\centering
\caption{Derived H$\alpha$ rotation curve of NGC~247, when considering both sides of the galaxy. Here, the large error bars do not represent the intrinsic uncertainty associated with the velocities. Instead, they reflect the large discrepancy arising between the solutions found for the rotation curve of the whole galaxy and that of the approaching side, the latter being affected by important non-circular motions.\label{a2_t2}}
\resizebox{0.46\textwidth}{!}{
\begin{tabular}{@{}cccccc@{}}
\hline
\hline
& & & & & \\
\textit{radius} & $V_{\rm rot}$ & ${\Delta}V_{\rm rot}$ & \textit{radius} & $V_{\rm rot}$ & ${\Delta}V_{\rm rot}$ \\
(arcsec) & (km s$^{-1}$) & (km s$^{-1}$) & (arcsec) & (km s$^{-1}$) & (km s$^{-1}$) \\
\hline
& & & & & \\
30 & 11.3 & 6.3 & 360 & 55.0 & 43.3 \\
60 & 46.3 & 10.0 & 390 & 63.7 & 30.2 \\
90 & 50.5 & 3.0 & 420 & 70.8 & 28.9 \\
120 & 51.5 & 5.7 & 450 & 80.4 & 6.7 \\
150 & 63.7 & 12.6 & 480 & 89.0 & 10.4 \\
180 & 74.2 & 7.9 & 510 & 98.0 & 4.3 \\
210 & 62.6 & 12.5 & 540 & 104.7 & 1.9 \\
240 & 65.4 & 26.4 & 570 & 109.3 & 0.3 \\
270 & 57.8 & 19.2 & 600 & 110.9 & 1.2 \\
300 & 52.3 & 25.9 & 630 & 93.6 & 5.3  \\
330 & 49.3 & 32.9 & & & \\
\hline
\end{tabular}}

\end{table}

\begin{table}
\centering
\caption{Adopted rotation curve for NGC~247, including the detections of faint H$\alpha$ emission associated with the radial intervals at $r=11.2'$ and $r=13.1'$ (bold font). The error bars are taken as those derived by the {\sc rotcur} task, except for those associated with the areas of extended emission (see Section 4). \label{a2_t3}}
\resizebox{0.46\textwidth}{!}{
\begin{tabular}{@{}cccccc@{}}
\hline
\hline
& & & & & \\
\textit{radius} & $V_{\rm rot}$ & ${\Delta}V_{\rm rot}$ & \textit{radius} & $V_{\rm rot}$ & ${\Delta}V_{\rm rot}$ \\
(arcsec) & (km s$^{-1}$) & (km s$^{-1}$) & (arcsec) & (km s$^{-1}$) & (km s$^{-1}$) \\
\hline
& & & & & \\
30 & 17.5 & 1.4 & 360 & 85.6 & 0.8 \\
60 & 55.7 & 1.1 & 390 & 93.8 & 0.7 \\
90 & 47.6 & 1.9 & 420 & 99.7 & 0.5 \\
120 & 45.8 & 3.3 & 450 & 87.1 & 2.5 \\
150 & 51.1 & 1.8 & 480 & 78.6 & 2.5 \\
180 & 74.4 & 1.6 & 510 & 93.7 & 1.7 \\
210 & 75.0 & 0.8 & 540 & 102.8 & 1.4 \\
240 & 76.9 & 0.6 &  &  &  \\
270 & 77.0 & 1.0 &  &  &  \\
300 & 71.8 & 1.9 & ${\bf 674.5\pm59.0}$ & {\bf 95.0} & {\bf 14.7} \\
330 & 82.1 & 1.2 & ${\bf 786.3\pm62.8}$ & {\bf 101.3} & {\bf 24.1} \\
\hline
\end{tabular}}

\end{table}

\section{\large{Mass models}}

Observational data seem to favour flat core mass models rather than cuspy core models \citep[see][]{Bla2001121,deB2002385,deB2003340,Kas2006162,Spa2008383}. We therefore choose to model a dark halo in the form of an isothermal sphere, which has a density profile as defined by Eq. \ref{a2_eq1}. This model is parameterized by a central density $\rho_{\rm 0}$ and core radius $r_{\rm c}$. The parameters are tied together by $\rho_{\rm 0}=9\sigma^2/4\pi$G${r_c}^2$, where $\sigma$ is the one dimensional velocity dispersion.
\begin{eqnarray}
\rho(r) = \frac{\rho_{\rm 0}}{[1 + (r/r_{\rm c})^2]^{3/2}}
\label{a2_eq1}
\end{eqnarray}

Based on a least-square fitting procedure, three parameters must be determined during the mass model analysis \citep[][]{Car1985294,Car1985299}. The first two are $\rho_{\rm 0}$ and $r_{\rm c}$, which characterize the dark halo, and the third is the mass-to-light ratio (\textit{M}/\textit{L})$_\star$, which characterizes the luminous component. We use the same surface brightness profile (\textit{B} band) as in \citet{Puc1990100} for NGC~300 and \citet{Car1990100} for NGC~247 to describe the stellar disc. The absolute magnitude of the sun is set to 5.43 in this band, and we take the same profile for the gaseous component as the one in the H\thinspace{\sc i} papers, which includes a 4/3 correction factor to account for helium. 

\begin{table}
\centering
\caption{Adopted H$\alpha$ rotation curve for NGC~300. The error bars are derived as the largest difference between the solutions found for the rotation curve of the whole galaxy and that of the approaching side or receding side, or if larger, the intrinsic error associated with the {\sc rotcur} task. \label{a2_t4}}
\resizebox{0.46\textwidth}{!}{%
\begin{tabular}{@{}cccccc@{}}
\hline
\hline
& & & & & \\
\textit{radius} & $V_{\rm rot}$ & ${\Delta}V_{\rm rot}$ & \textit{radius} & $V_{\rm rot}$ & ${\Delta}V_{\rm rot}$ \\
(arcsec) & (km s$^{-1}$) & (km s$^{-1}$) & (arcsec) & (km s$^{-1}$) & (km s$^{-1}$) \\
\hline
& & & & & \\
45 & 12.8 & 2.4 & 450 & 80.6 & 1.2 \\
90 & 27.1 & 3.8 & 495 & 77.1 & 18.8  \\
135 & 43.0 & 3.0 & 540 & 73.6 & 5.4 \\
180 & 47.5 & 15.1 & 585 & 67.2 & 4.5 \\
225 & 54.4 & 12.6 & 630 & 68.9 & 3.7 \\
270 & 64.6 & 14.4 & 675 & 66.9 & 11.6 \\
315 & 66.5 & 11.4 & 720 & 73.0 & 1.8 \\
360 & 74.1 & 7.0 & 765 & 67.5 & 2.3 \\
405 & 73.9 & 1.3 & 810 & 86.1 & 2.0 \\
\hline
%\hline
\end{tabular}}
\end{table}
\begin{figure}
\centering
\begin{minipage}[c]{0.9\linewidth}
  \centering \includegraphics[width=\linewidth]{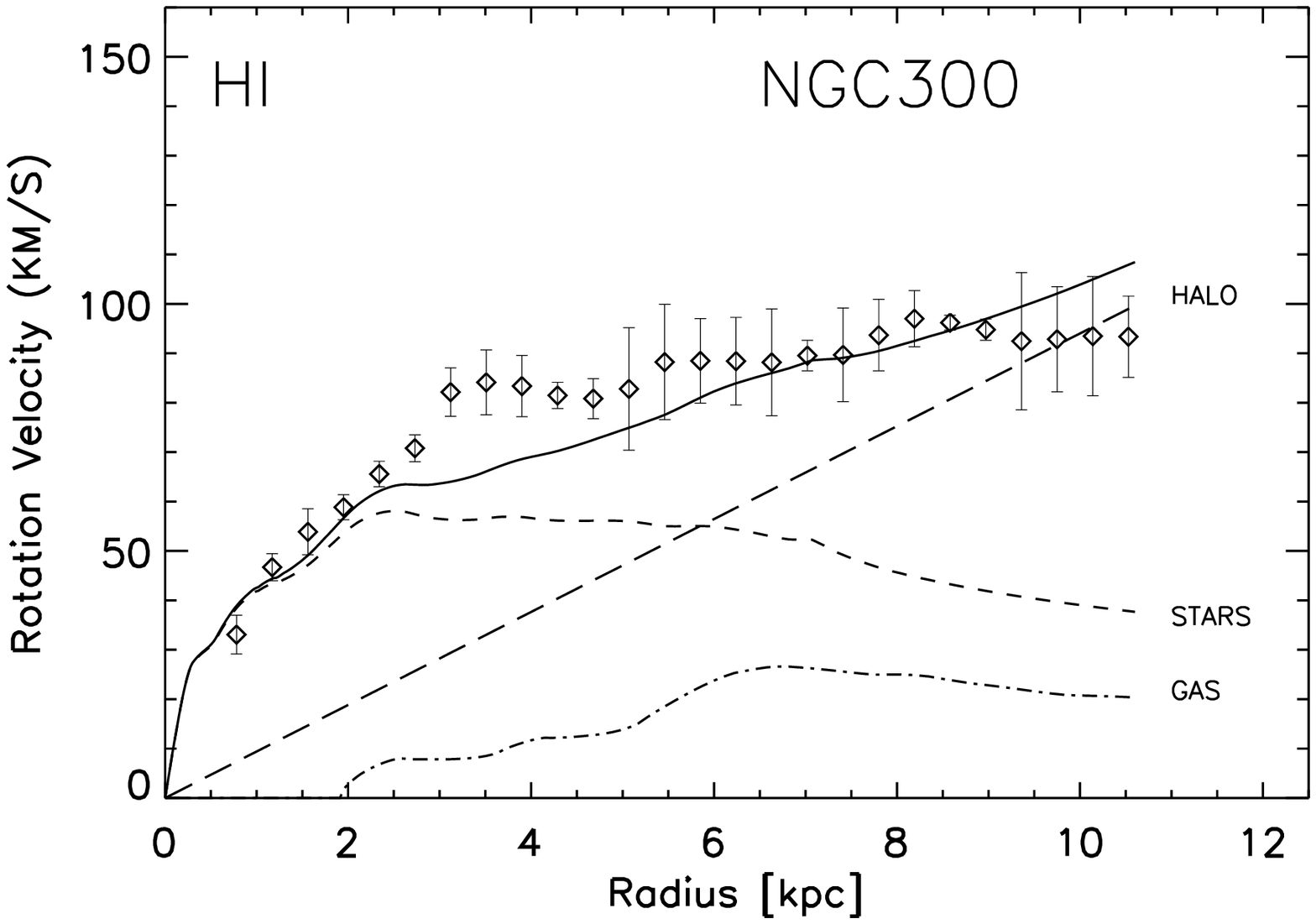}
\end{minipage}
\begin{minipage}[c]{0.9\linewidth}
  \centering \includegraphics[width=\linewidth]{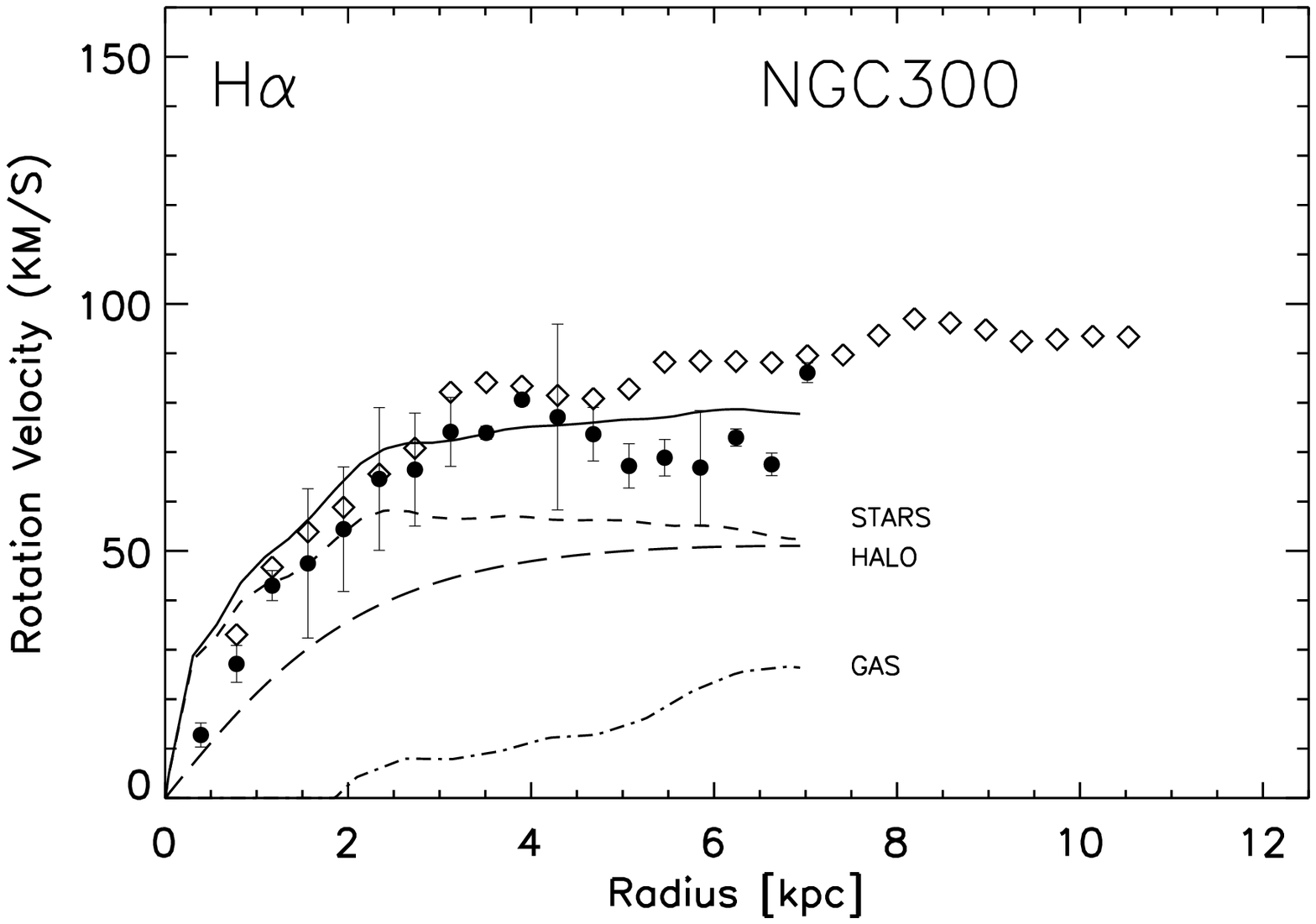}
\end{minipage}
\begin{minipage}[c]{0.9\linewidth}
  \centering \includegraphics[width=\linewidth]{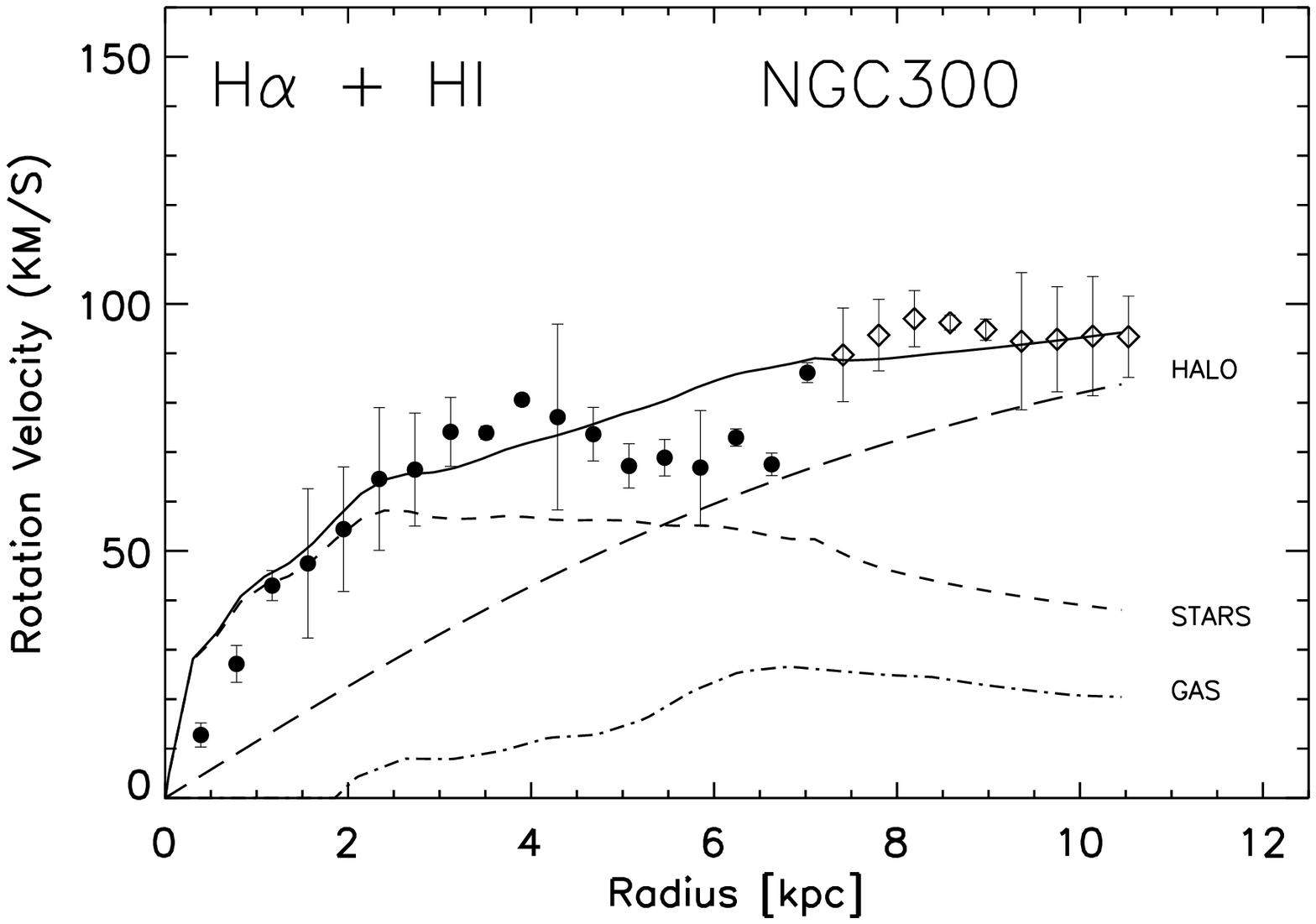}
\end{minipage}
\caption{Maximum stellar disc mass models for NGC~300, where an isothermal sphere was used for the dark halo. The stellar disc is plotted with a small-dashed line, the gaseous component with a dot-dashed line, the halo with a long-dashed line and the total with the continuous line. Top: Mass model for the adopted H\thinspace{\sc i} rotation curve (diamond symbols) taken from \citet{Puc1990100}. Middle: Same but for the H$\alpha$ rotation curve derived in this study (filled black circles). H\thinspace{\sc i} data are also shown with diamond-shape points. Bottom: Same but for the combined H$\alpha$ ($r<13.6'$ or $r<7.1$ kpc) + H\thinspace{\sc i} ($r>13.6'$ or $r>7.1$ kpc) rotation curve. The parameters extracted are in Table \ref{a2_t5}. }
\label{a2f8a}
\end{figure}
\begin{figure}
\centering
\begin{minipage}[c]{0.9\linewidth}
  \centering \includegraphics[width=\linewidth]{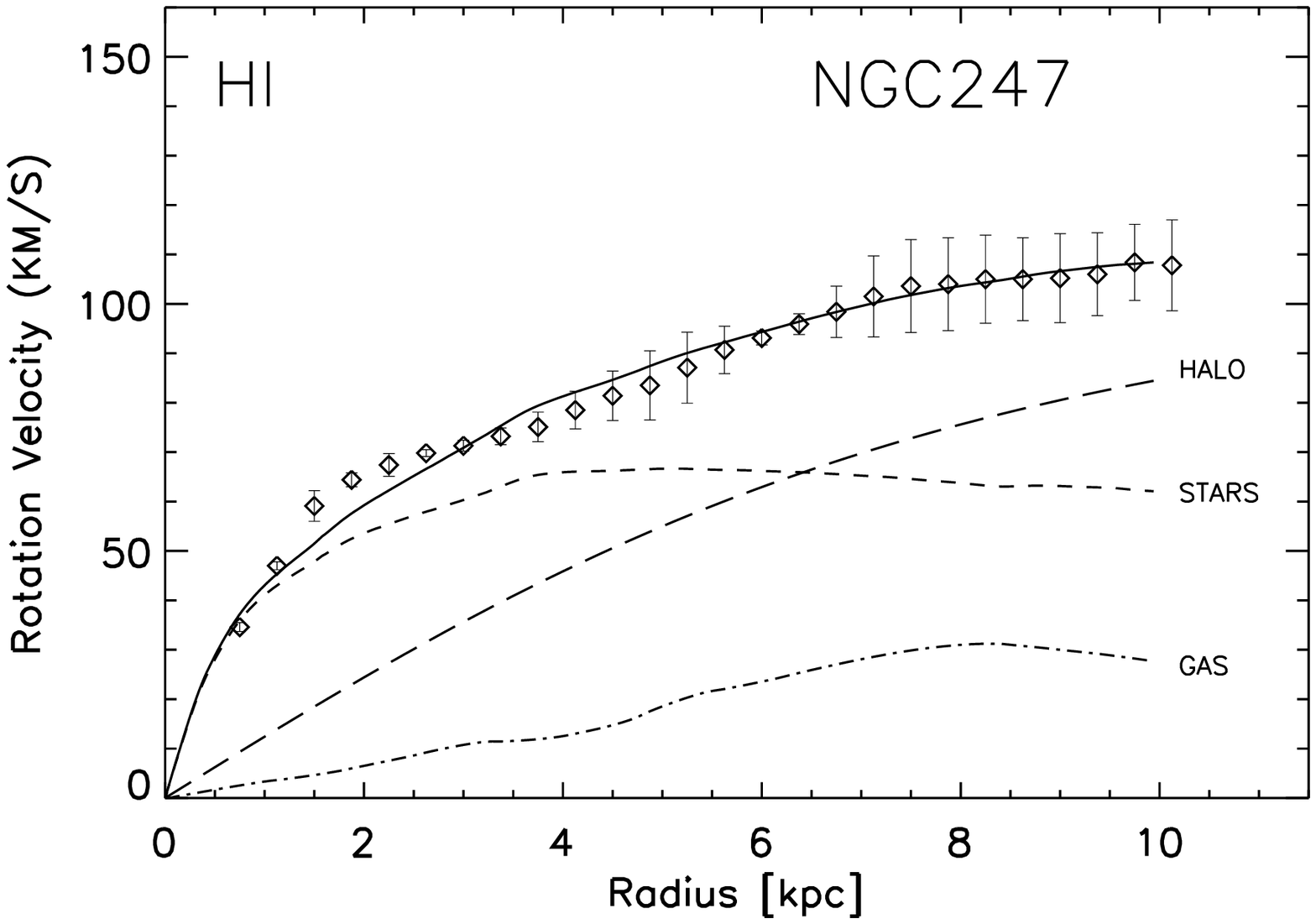}
\end{minipage}
\begin{minipage}[c]{0.9\linewidth}
  \centering \includegraphics[width=\linewidth]{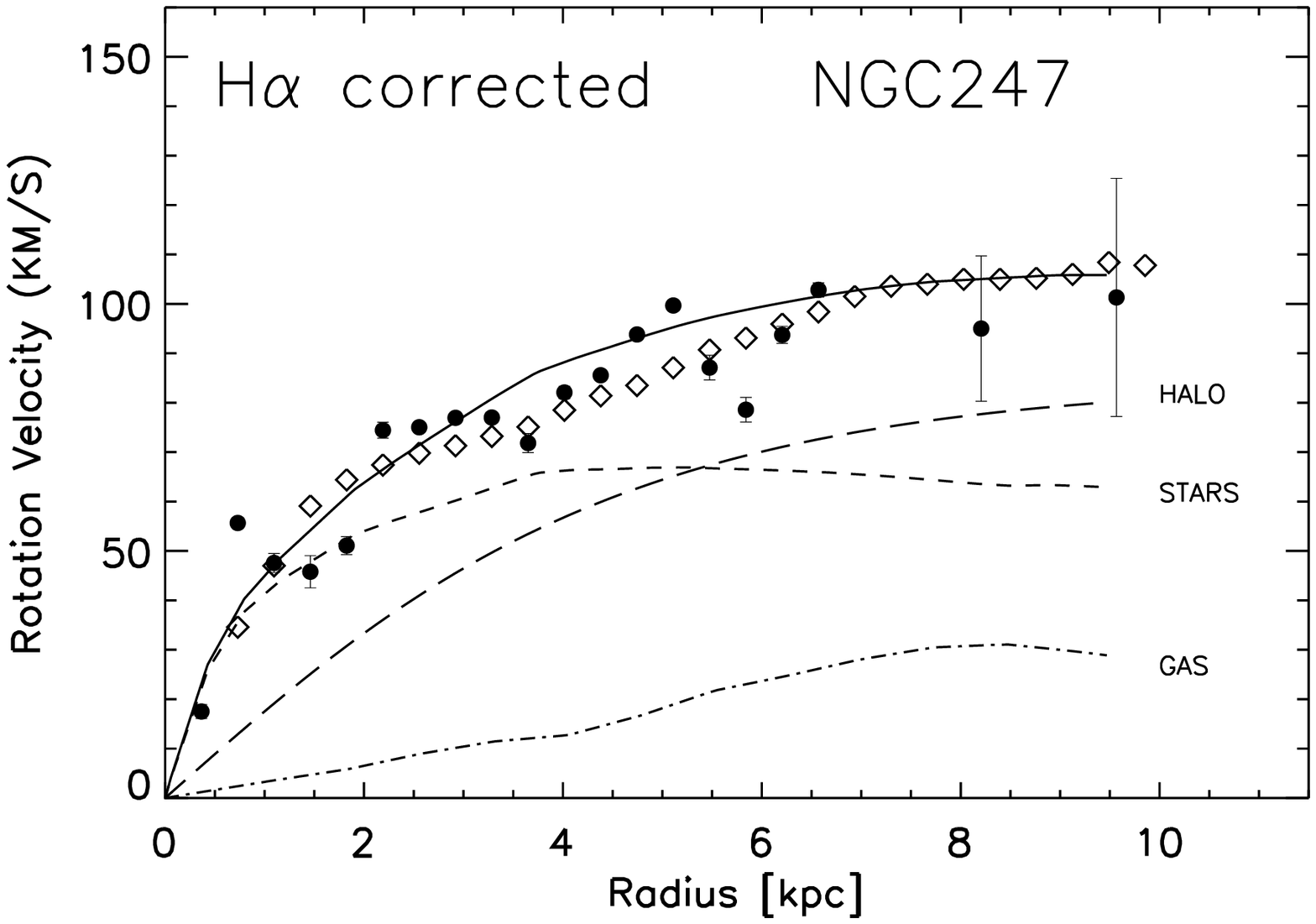}
\end{minipage}
\caption{Same caption as Fig. \ref{a2f8a}, but for NGC~247. Here, we analysed the H\thinspace{\sc i} (top) and the H$\alpha$ (bottom) rotation curve derived for the receding side (not affected by the non-circular motions), which includes the two detections of diffuse emission at large radii. The first is shown with diamond-like symbols, and the second is shown with the circles. The parameters extracted are in Table \ref{a2_t6}. }
\label{a2f8b}
\end{figure}

Best-fitting models generally reproduce well the rotation curves. However, various combinations of parameters can give the same results, {\rm i. e.} these models are degenerate \citep{vanA1986320, Car1985294, Dut2005619}. Models such as those with no dark halo or those that are sub-maximum can be more interesting since they give only one solution. We applied a model with no dark halo, but this gave very poor results for both galaxies. 
 
Instead, we choose to use the model developed by \citet[][]{Cou1999513}. These authors argued that for the majority of high surface brightness (HSB) late-type galaxies, pure maximum stellar discs were not adequate. They then observed that the Tully-Fisher relation did not seem to show any dependence on surface brightness for HSB galaxies when plotted as \textit{M}$_{\rm r}$ versus \textit{V}$_{2.2}$ (velocity at 2.2 exponential scale lengths), and using this, they modelled the contraction of dark haloes by adiabatic infall of baryons. Their models then suggested that discs were sub-maximum with $v_{\rm disc}/v_{\rm total}\sim0.6$, which requires that (\textit{M}/\textit{L})$_\star$ be fixed at a specific value. For NGC~300, ($M/L$)$_\star$ must therefore be $\sim1.8M_\odot/L_\odot$ and for NGC~247, $\sim4.9M_\odot/L_\odot$. For these particular galaxies, the ratios are consistent with the models being maximum-discs. 

For NGC~300, we analysed a total of three rotation curves (see Fig. \ref{a2f8a} and Table \ref{a2_t5}): first, the H\thinspace{\sc i} curve from \citet{Puc1990100}, second, the H$\alpha$ curve derived here, and third, the combined H$\alpha$ ($r<13.6'$) and H\thinspace{\sc i} ($r>13.6'$) curve, since the latter extends further out\footnote[4]{Since the $i$ and PA are compatible between studies, we can combine both rotation curves.}. For NGC~247 (see Fig. \ref{a2f8b} and Table \ref{a2_t6}), only the H$\alpha$ and H\thinspace{\sc i} curves were analysed since our H$\alpha$ rotation curve extends to the end of the H\thinspace{\sc i} disc. For the H\thinspace{\sc i} curves, we include the error bars derived in \citet{Puc1990100} for NGC~300 and in \citet{Car1990100} for NGC~247, and for our H$\alpha$ curves, we include the error bars associated with the adopted rotation curves (see Table 4 and Table 3, respectively).

\section{\large{Discussion}}\label{a1c6}

We have obtained very deep H$\alpha$ observations of NGC~247 and NGC~300, in order to study the extent of the optical disc in these galaxies. For NGC~300, we detected emission all the way to the extremities of our field of view ($24\times24'$), but suspect that there is still diffuse emission beyond. Other pointings and/or a larger field of view would be necessary to study the total extent of ionized gas in NGC~300. In the first section, we briefly discuss the results concerning this galaxy. We then address in greater detail the case of NGC~247, where we were able to detect H$\alpha$ emission on the entirety of the H\thinspace{\sc i} disc. 

\subsection{\normalsize NGC~300}\label{a1c61}

NGC~300 is a late-type spiral galaxy with a very rich distribution of bright H\thinspace{\sc ii} regions \citep[][]{Riz2006234}, and a very extended stellar disc reaching $r=24'$ that shows no sign of truncation \citep{Bla2005629}. The faintest level of the stellar disc reaches a surface brightness of 32 mag arcsec$^{-2}$ (in $V$ and $B$ band), and the stellar population was found to be mostly old \citep{Vla2009697}. The H\thinspace{\sc i} disc extends to $r\sim20'$ \citep{Puc1990100}, and seems to be flat beyond $r=15'$ (8 kpc). 
\begin{table}
\centering
\caption{Mass models for NGC~300 \label{a2_t5}}
\begin{minipage}[t]{14cm}
    \renewcommand{\footnoterule}{}
\begin{tabular}{@{}llll@{}}
\hline
\hline
Rotation Curve & H\thinspace{\sc i}\footnote[1]{Puche et al. (1990)} & 36-cm H$\alpha$ & H\thinspace{\sc i}+36-cm H$\alpha$  \\
\hline
\textbf{Sub-Max} & & & \\
(\textit{M}/\textit{L}$_B$)$\star$& 1.8 & 1.8 & 1.8 \\
& & & \\
Dark halo\footnote[2]{$r_{\rm c}$ in kpc, $\sigma$ in km s$^{-1}$ and $\rho_{\rm 0}$ in 10$^{-3}$ $M_\odot$ pc$^{-3}$} & $r_{\rm c}=173.6$ & $r_{\rm c}=2.4$ & $r_{\rm c}=9.7$ \\ 
% & $\sigma=934$ & $\sigma=$  & $\sigma=$ \\
& $\rho_{\rm 0}=4.9$ & $\rho_{\rm 0}=28.8$ & $\rho_{\rm 0}=7.4$ \\
\hline
\end{tabular}
\end{minipage}
\end{table}

We have obtained very deep H$\alpha$ observations of this galaxy. We reach EMs on the order of $0.1$ cm$^{-6}$ pc, and our rotation curve extends up to $r=13.5'$. These are to our knowledge the deepest H$\alpha$ observations of this galaxy. Other H$\alpha$ studies include that of \citet{Mar1985151} and \citet{Hoo1996112}. The first obtained a rotation curve extending only to $r=10'$, and the second reached EMs between 6 and 500 cm$^{-6}$ pc. We detect significantly more diffuse emission. Fig. \ref{a2f4} (bottom-left panel) shows that the bright H\thinspace{\sc ii} regions are embedded in a large diffuse background. \citet{Hoo1996112} estimated that the diffuse fraction of H$\alpha$ gas after scattering correction lies between 44 to 54 per cent of the total H$\alpha$ emission. Our study, which finds more diffuse gas, suggests that this fraction could be significantly larger. Since we were limited by our field of view, we suspect that there is more diffuse gas beyond our maximum radius ($r=13.5'$).    

\citet{Puc1990100} found a distinct spiral structure on the southern-east side of the galaxy at $r\sim20'$. It would be very interesting to probe this structure, as well as to see if the H$\alpha$ disc reaches the end of the H\thinspace{\sc i} disc, as it has been the case for other Sculptor group galaxies \nocite{Dic2008135,Bla1997490, Hla2010}(Dicaire et al. 2008 for NGC~7793; Bland-Hawthorn et al. 1997 and Hlavacek-Larrondo et al. 2010 for NGC~253; and this study for NGC~247). 
\begin{table}
\centering
\caption{Mass models for NGC~247 \label{a2_t6}}
\begin{minipage}[t]{14cm}
    \renewcommand{\footnoterule}{}
\begin{tabular}{@{}lll@{}}
\hline
\hline
Rotation Curve & H\thinspace{\sc i}\footnote[1]{\citet{Car1990100}} & 36-cm H$\alpha$ (adopted curve) \\
\hline
\textbf{Sub-Max} & & \\
(\textit{M}/\textit{L}$_B$)$\star$& 4.9 & 4.9 \\
& & \\
Dark halo\footnote[2]{$r_{\rm c}$ in kpc, $\sigma$ in km s$^{-1}$ and $\rho_{\rm 0}$ in 10$^{-3}$ $M_\odot$ pc$^{-3}$} & $r_{\rm c}=8.8$ & $r_{\rm c}=5.0$ \\ 
% & $\sigma=62.9$ & $\sigma=$ \\
& $\rho_{\rm 0}=8.6$ & $\rho_{\rm 0}=17.7$ \\
\hline
\end{tabular}
\end{minipage}
\end{table}

In Fig. \ref{a2f7}, we show the kinematical parameters and rotation curve we derive for NGC~300. Although our H$\alpha$ data have a slightly better spatial resolution than the H\thinspace{\sc i} data ($2.8''\times2.8''$ compared to $5''\times5''$ for the central and $17''\times12''$ for the outer parts in H\thinspace{\sc i}), they could still be affected by beam smearing in the inner regions. Both rotation curves are nonetheless consistent, and the only significant difference is seen at large radii ($r=10-14'$) where the H$\alpha$ rotation curve appears to drop and then raise again. This is seen in both the receding and approaching sides. The variation of the kinematical parameters with radius shows that there could be a warp in the disc starting at $r\sim9'$. This is suggested by the continuous increase of the PA with radius starting at $r\sim9'$, also seen in the H\thinspace{\sc i} data \citep{Puc1990100}, meaning that the warp seems to affect both the H\thinspace{\sc i} and optical disc. The PV diagram is shown in the bottom-panel of Fig. \ref{a2f4}. Here, we see that the rotation curve we derive seems to coincide with the maximum intensity at a given radius, indicating that the velocities we derive are able to reproduce the velocity structure in NGC~300. This diagram also confirms that the dynamical centre is offset by $\sim30''$ from the photometric one, which can be seen in the PV diagram as continuum emission at an angular offset of $-30''$.

We now briefly address the results concerning the mass models (see Fig. \ref{a2f8a} and Table \ref{a2_t5}). We applied a model which required ($M/L$)$_\star=1.8M_\odot/L_\odot$. This ($M/L$)$_\star$ is in good agreement with the one predicted by stellar population models. \citet{Bel2001550} predict for NGC~300, using a (\textit{B}-\textit{V}) magnitude of 0.59 mag given by the RC3 catalogue, that (\textit{M}/\textit{L})$_\star$ in the \textit{B} band is 1.2 M$_\odot$/L$_\odot$. ($M/L$)$_\star=1.8M_\odot/L_\odot$ is also consistent with the value required to maximise the contribution of the stellar disc in the inner regions (i. e. a maximum stellar disc model). However, this value of ($M/L$)$_\star$, and even the one predicted by stellar population models, seems to slightly overestimate the inner parts of the rotation curve, at least for two H$\alpha$ velocity points.  

Fig. \ref{a2f8a} shows that the maximum disc model reproduces well all three rotation curves, but the parameters derived for the H$\alpha$ and combined H$\alpha$ and H\thinspace{\sc i} rotation curves are significantly different than those of the H\thinspace{\sc i} rotation curve alone. Our results suggest that the first two (H$\alpha$ and H$\alpha$+H\thinspace{\sc i}) have a smaller core radius, but this can be explained by the slightly lower velocities observed in H$\alpha$ between $r=5-7$ kpc ($r=10-14'$). It seems that since the stellar component reproduces well the rotation curve at $r<3$ kpc, the mass model then tries to adjust the dark halo component to reproduce the shape of the curve for $r>3$ kpc. Between $r=5-7$ kpc, the H$\alpha$ rotation curve has lower velocities, which could make the model converge towards a more concentrated halo (smaller and more realistic $r_{\rm c}$). 

\begin{figure}
\centering
\begin{minipage}[c]{0.6\linewidth}
 \hspace{0.4cm} \includegraphics[width=\linewidth]{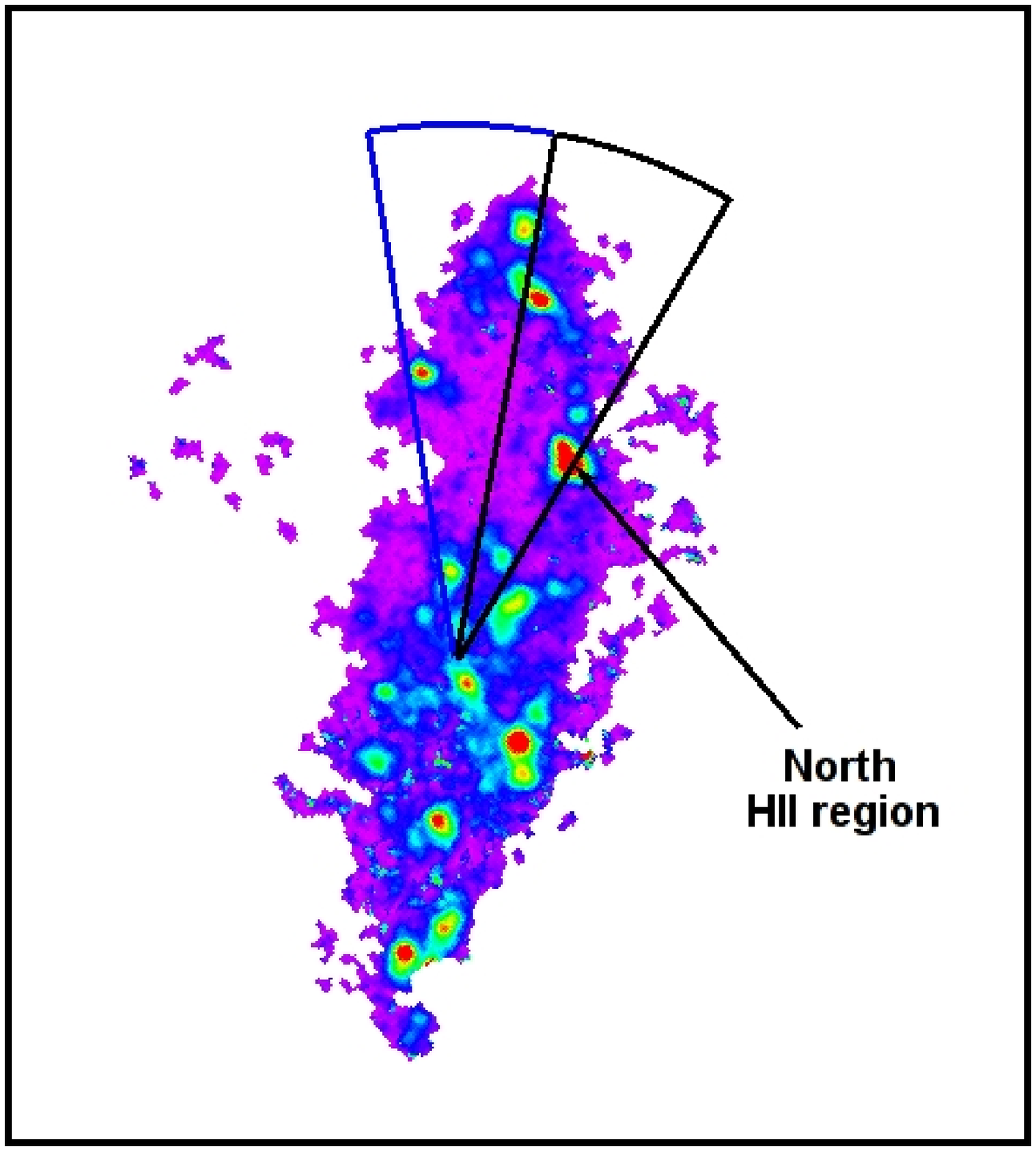}
\end{minipage}
\begin{minipage}[c]{0.99\linewidth}
  \centering \includegraphics[width=\linewidth]{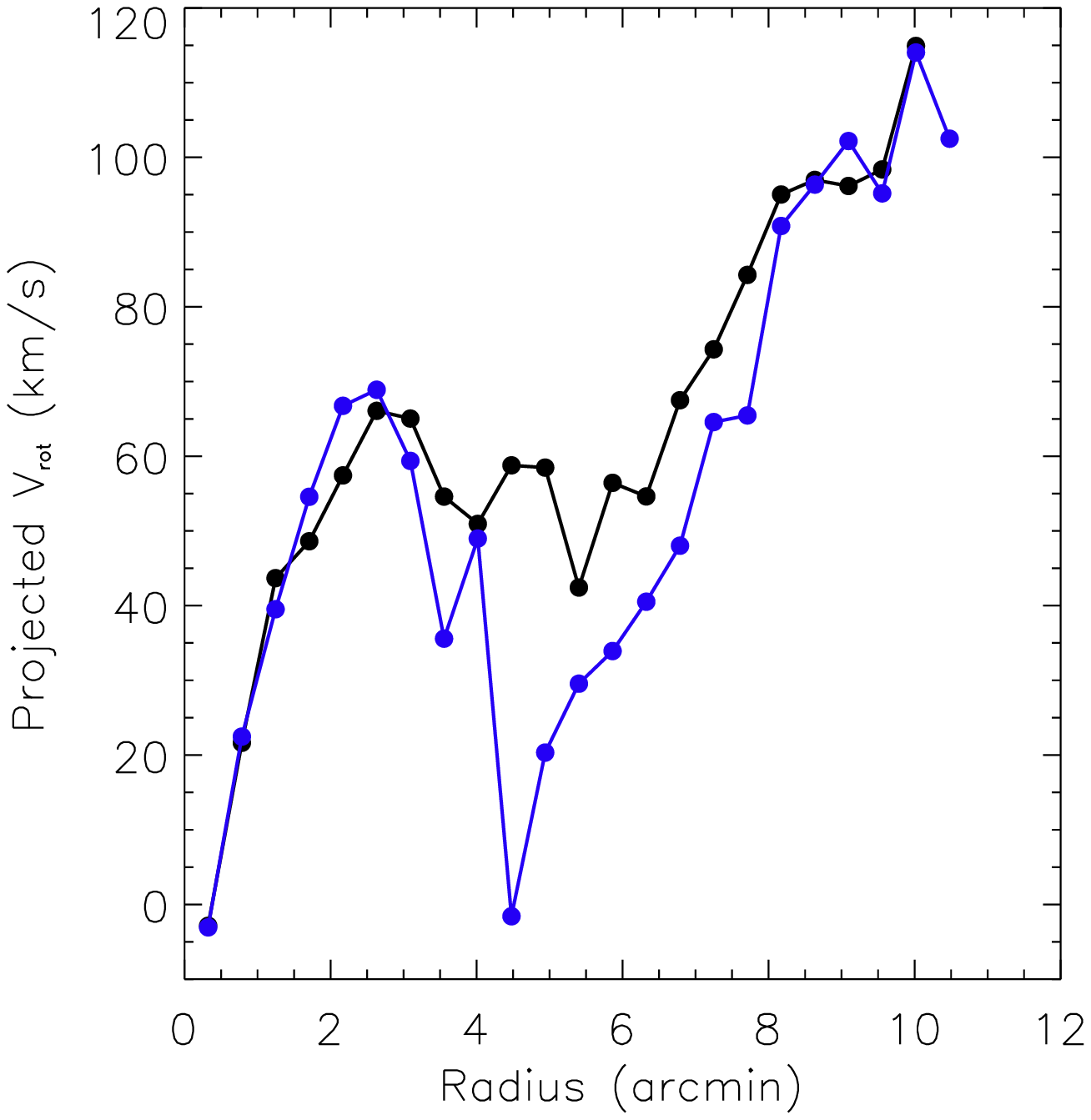}
\end{minipage}
\caption{Flux calibrated integrated map of NGC~247 shown in the top panel. The bright H\thinspace{\sc ii} region to the north-west was believed to be an H\thinspace{\sc ii} ring causing a distortion in the rotation curve. We extract two regions along the north oppositely symetric with respect to the major axis (shown with the blue and black contours), and plot in the bottom panel the derived projected rotation velocity as a function of radius. By projected rotation velocity, we mean that the observed radial velocity has been corrected for $V_{\rm sys}$ and $i$, but not for the cosine dependence on the azimuthal angle in the plane of the galaxy. The region within the blue contours contains most of the diffuse emission of the interarm region, and the region within the black contours contains the north H\thinspace{\sc ii} region and some diffuse emission of the interarm region. The bottom panel shows that the region dominated by the diffuse interarm emission has lower rotation velocities than when excluding this region. }
\label{fig10}
\end{figure}
\subsection{\normalsize NGC~247}\label{a1c62}

\subsubsection{Detection of extended H$\alpha$ emission}

In NGC~247, we could study the extent of the H$\alpha$ emission on the entirety of the H\thinspace{\sc i} disc. Our observations reach EMs of 0.1 cm$^{-6}$ pc (not including the fainter emission detected using the annular ring binning in Section 3.1). This makes our study one of the deepest H$\alpha$ studies of NGC~247. \citet{Fer1996111} looked at diffuse ionized gas in NGC~247, but only reached a level of 6 cm$^{-6}$ pc for H$\alpha$+[N\thinspace{\sc ii}], and a maximum radius of $r=10'$. We detect much more diffuse emission, especially between the spiral arms (although still quite faint and with a disturbed velocity field in the northern arm). We also find emission up to $r=13.1'$. \citet{Fer1996111} determined that diffuse gas contributes to $\sim50$ per cent of the total H$\alpha$ luminosity in NGC~247. We suspect, as for NGC~300, that this fraction is much higher if we include all of the fainter gas we detect. This emphasises that diffuse ionized gas can contribute significantly in a galaxy, and that it can be used to probe its outer disc.  

In \citet[][]{Hla2010}, we succeeded in detecting [N\thinspace{\sc ii}] ($\lambda$ 6548 \AA) emission \textit{beyond the H\thinspace{\sc i} disc} in NGC~253, confirming the declining part of the rotation curve. \citet{Bla1997490} reported that the [N\thinspace{\sc ii}]($\lambda$ 6548 \AA)/H$\alpha$ ratio in NGC~253 was quite strong ($\sim$1) beyond the H\thinspace{\sc i} disc, and so our detection of [N\thinspace{\sc ii}] was not surprising. \citet{Ken2008178} find a [N\thinspace{\sc ii}]($\lambda$ 6548 \AA\ and $\lambda$ 6583 \AA)/H$\alpha$ ratio of 0.48 for NGC~253, but only of 0.24 for NGC~247, suggesting an even smaller fraction for [N\thinspace{\sc ii}]($\lambda$ 6548 \AA)/H$\alpha$ and making it more difficult to detect this line in NGC~247.  

By binning the data in annular rings, we were able to detect extended faint H$\alpha$ emission on the northern side of NGC~247 at $r=11.2'$ and $r=13.1'$. These two detections have respectively an EM of $\sim0.02$ cm$^{-6}$ pc and $\sim0.05$ cm$^{-6}$ pc. We estimate that these values are within a factor of two accurate, given that the mean absolute deviation of the difference between the annular ring spectrum and the contaminating reflection spectrum has a noise level of about 40 per cent. Based on this estimate of the noise level, our detection at $r=11.2'$ represents a $3\sigma$ detection and the one at $r=13.1'$ represents at least a $2\sigma$ detection. We also only detect H$\alpha$ emission up to $r=8.5'$ on the southern side of the galaxy. Although it does seem that the northern side of NGC~247 has a larger amount of gas compared to the southern side \citep[see {\rm e. g.}][]{Car1990100}, our observations could be biased because the northern side of the galaxy has typical radial velocities of 80 km s$^{-1}$, corresponding to a filter transmission of 80 per cent (maximum value for our filter), whereas the southern side has radial velocities of 250 km s$^{-1}$, corresponding to a filter transmission probability of only 65 per cent. However, one of the main conclusions of our observations is that, unlike NGC~253, we do not detect any optical emission beyond the H\thinspace{\sc i} disc. We now examine the implications of this non detection in NGC~247. 

\subsubsection{Implications of the extended H$\alpha$ emission}

\citet{Bla1997490} proposed several explanations as to why NGC~253 had an optical disc larger than the H\thinspace{\sc i} disc. Among them, they first looked at the metagalactic ionizing background, which predicted an upper limit EM of $0.25$ cm$^{-6}$ pc, but their H$\alpha$ observations had EMs much higher. They then proposed photoionization models (Lyman continuum, Lyc, ionizing photons), which predicted a much lower fraction of [N\thinspace{\sc ii}]/H$\alpha$ beyond the H\thinspace{\sc i} disc than their data suggested. Finally, they proposed ionizing photons from hot young stars located near the central regions of the galaxy, which could look out towards the outer disc through the galaxy's warp. Through dust scattering, these photons would have to travel through less dense environments, i.e. they would be less absorbed, and could ionize the gas beyond the H\thinspace{\sc i} disc. Our [N\thinspace{\sc ii}] and H$\alpha$ observations of NGC~253 \citep{Hla2010} favoured this explanation. 

If this theory holds, this would mean that galaxies without a warp could \emph{not} ionize gas beyond the H\thinspace{\sc i} disc. NGC~247 is not warped, and we do not detect H$\alpha$ emission beyond the H\thinspace{\sc i} disc. Hence, our results are consistent with this scenario.

Our H$\alpha$ measurements for NGC~247 fall below the predicted upper limit of the metagalactic ionizing background. However, this theory predicts that there should be optical emission beyond the H\thinspace{\sc i} disc, which we do not observe. As for the Lyc ionizing photons, UV GALEX maps of NGC~247 show diffuse emission extending up to $\sim$ 11-12$'$, which coincides with the first of the two annular rings where diffuse gas was detected. Since there is no significant UV emission beyond $r=13'$, this could also explain why we did not see optical emission beyond the H\thinspace{\sc i} disc. However, photoionization models still have many problems relating to the details of how these photons are able to escape to large distances. The exact ionizing source of diffuse ionized gas still remains unclear \citep[see a recent review on the topic by][]{Haf200981}, but our \emph{non detection} beyond the H\thinspace{\sc i} disc in NGC~247 and \emph{detection} beyond the H\thinspace{\sc i} disc for the warped galaxy NGC~253 seem to favour the scenario in which photons are only able to leak out towards the outer disc through a warp.

\subsubsection{Kinematical analysis}

We show in Fig. \ref{a2f3}, the velocity field, dispersion map and integrated map of NGC~247. Fig. 6 then shows that the rotation curve derived for the approaching (northern) side has a large drop in velocities at $r\sim5'$. This feature was also seen in the H\thinspace{\sc i} study by \citet{Car1990100}, as well as the optical study by \citet{Car1985294}, and was interpreted as being caused by non-circular motions associated with a bright H\thinspace{\sc ii} ring to the north-west.

In Fig. 10, we explore in more detail this feature. The bright H\thinspace{\sc ii} region to the north-west at $r\sim5'$ is believed to be the ring causing the shift in velocities. We extract two regions on the approaching side of the galaxy, oppositely symmetric with respect to the major axis. These are shown with the blue and black contours. The region within the blue contours contains most of the diffuse emission of the interarm region, and the region within the black contours contains the H\thinspace{\sc ii} region and some diffuse emission of the interarm region. In the bottom panel, we plot the derived projected rotation velocity as a function of radius for each region, which is the observed radial velocity corrected for the systemic velocity and inclination of the galaxy, but not for the cosine dependence on the azimuthal angle in the plane of the galaxy. Since both regions are symmetric with respect to the major axis, they will both be affected in the same way by the azimuthal angle dependence. The derived projected velocities can therefore be directly compared with one another. 

The bottom panel of Fig. 10 shows that the region dominated by the diffuse interarm emission (in blue) has much lower velocities. Hence, the H\thinspace{\sc ii} region to the north is not the cause of the drop in the rotation curve, which is not in agreement with the interpretation of \citet{Car1990100}. Instead, it seems that the diffuse region between the spiral arms to the north is causing the lower rotation velocities seen between $r=4-7'$. The velocity field of the interarm region has a distorted pattern with radial velocities varying between $50-250$ km s$^{-1}$, suggesting that this region contains important non-circular motions. This is also supported by the PV diagram in Fig. \ref{a2f3}, which shows diffuse emission at $r\sim4'$ (or $r\sim240-250''$) of large velocities that do not coincide with the maximum intensities.

As mentioned earlier, in the PV diagram of Fig. \ref{a2f3}, we show in black the rotation curve we derive for both sides, which is affected by the non-circular motions of the interarm region. This curve falls below the maximum intensities on the receding side of the galaxy, consistent with the idea that it does not represent the true rotation within the galaxy. In red, we show the rotation curve derived only using the receding side, which follows well the maximum intensities in the PV diagram, and therefore represents well the true rotation. The latter was used when deriving the mass models. 

Fig. \ref{a2f8b} and Table \ref{a2_t6} show the solutions derived for the mass model analysis. The model required a (\textit{M}/\textit{L})$_\star$ ratio of 4.9 M$_\odot$/L$_\odot$, and gave a good fit for both rotation curves. The (\textit{M}/\textit{L})$_\star$ ratio is quite high when compared to stellar population models \citep[{\rm e. g.} ][]{Bel2001550}, which predict a ratio of 1.0 M$_\odot$/L$_\odot$ in the \textit{B} band, using a (\textit{B}-\textit{V}) magnitude of 0.56 mag as given by the RC3 catalogue. However, it is interesting to note that (\textit{M}/\textit{L})$_\star=4.9$ M$_\odot$/L$_\odot$ is again consistent with the value required to maximise the contribution of the stellar disc in the inner regions, just like in NGC~300. Our results show that the H$\alpha$ rotation curve requires a larger $\rho_{\rm 0}$ (halo is more concentrated). This could be because of the slightly higher H$\alpha$ velocity points between $r=3-7$ kpc, and the two extended points found with the annular binning method which have lower velocities than the H\thinspace{\sc i} rotation curve. The model could therefore be trying to find a more concentrated dark halo.

\section{\large{Concluding remarks}}

This paper is the third in a series that aims to study diffuse ionized gas of the Sculptor group galaxies using very deep optical observations. The first was carried out by \citet{Dic2008135} and confirmed that NGC~7793 had a truly declining rotation curve. The second looked at NGC~253 \citep{Hla2010} and found very extended [N\thinspace{\sc ii}] emission beyond the H\thinspace{\sc i} disc, showing a significant decline in the rotation curve. The third is presented here, and looks at H$\alpha$ emission in NGC~247 and NGC~300. For NGC~247, we have detected diffuse H$\alpha$ on the entirety of the H\thinspace{\sc i} disc, but not beyond. For NGC~300, H$\alpha$ emission is detected on the entirety of the allowed spatial coverage, but since we are limited by our field of view, we suspect that there is more optical emission beyond. However, our study still remains the deepest H$\alpha$ study of these two galaxies, and from there, we have obtained the most extended optical rotation curves thus far for both galaxies. To account for the existence of diffuse optical emission beyond the H\thinspace{\sc i} disc in the warped galaxy NGC~253, but \emph{not} in NGC~247 (which is not warped), our observations favour ionization by hot young stars in the central region of the galaxy, which only through a warp, can let photons escape and ionize the interstellar medium in the outer parts.

\section*{Acknowledgements}

We would like to thank the referee for taking the time to make his/her thorough review, and for providing very helpful comments, which have improved the paper greatly. We would also like to thank the staff from the ESO La Silla site, in Chile, for their support, as well as Monica Relano-Pastor for taking time from her busy schedule to pre-read this article. Moreover, we would also like to thank the students from the 1.2-m Euler Telescope for lending us their computer during the observation mission. We recognize all the support given by the Natural Sciences and Engineering Research Council of Canada, as well as the Fonds Qu\'{e}b\'{e}cois de la Recherche sur la Nature et les Technologies. The Southern H-Alpha Sky Survey Atlas (SHASSA), which is supported by the National Science Foundation, was used for flux calibration of the data in this study. The images were taken with a robotic camera operating at Cerro Tololo Inter-American Observatory (CTIO), in Chile.

\label{lastpage}

\bibliographystyle{mn2e}
\bibliography{247300bib}

\begin{thebibliography}{}

\bibitem[\protect\citeauthoryear{{Begeman}}{{Begeman}}{1989}]{Beg1989223}
{Begeman} K.~G.,  1989, \aap, 223, 47

\bibitem[\protect\citeauthoryear{{Bell} \& {de Jong}}{{Bell} \& {de
  Jong}}{2001}]{Bel2001550}
{Bell} E.~F.,  {de Jong} R.~S.,  2001, \apj, 550, 212

\bibitem[\protect\citeauthoryear{{Blais-Ouellette}, {Amram} \&
  {Carignan}}{{Blais-Ouellette} et~al.}{2001}]{Bla2001121}
{Blais-Ouellette} S.,  {Amram} P.,    {Carignan} C.,  2001, \aj, 121, 1952

\bibitem[\protect\citeauthoryear{{Bland-Hawthorn}, {Freeman} \&
  {Quinn}}{{Bland-Hawthorn} et~al.}{1997}]{Bla1997490}
{Bland-Hawthorn} J.,  {Freeman} K.~C.,    {Quinn} P.~J.,  1997, \apj, 490, 143

\bibitem[\protect\citeauthoryear{{Bland-Hawthorn}, {Vlaji{\'c}}, {Freeman} \&
  {Draine}}{{Bland-Hawthorn} et~al.}{2005}]{Bla2005629}
{Bland-Hawthorn} J.,  {Vlaji{\'c}} M.,  {Freeman} K.~C.,    {Draine} B.~T.,
  2005, \apj, 629, 239

\bibitem[\protect\citeauthoryear{{Bosma}}{{Bosma}}{1978}]{Bos1978}
{Bosma} A.,  1978, PhD thesis, PhD Thesis, Groningen Univ., (1978)

\bibitem[\protect\citeauthoryear{{Carignan}}{{Carignan}}{1985a}]{Car1985299}
{Carignan} C.,  1985a, \apj, 299, 59

\bibitem[\protect\citeauthoryear{{Carignan}}{{Carignan}}{1985b}]{Car198558}
{Carignan} C.,  1985b, \apjs, 58, 107

\bibitem[\protect\citeauthoryear{{Carignan} \& {Freeman}}{{Carignan} \&
  {Freeman}}{1985}]{Car1985294}
{Carignan} C.,  {Freeman} K.~C.,  1985, \apj, 294, 494

\bibitem[\protect\citeauthoryear{{Carignan} \& {Puche}}{{Carignan} \&
  {Puche}}{1990}]{Car1990100}
{Carignan} C.,  {Puche} D.,  1990, \aj, 100, 641

\bibitem[\protect\citeauthoryear{{Courteau} \& {Rix}}{{Courteau} \&
  {Rix}}{1999}]{Cou1999513}
{Courteau} S.,  {Rix} H.,  1999, \apj, 513, 561

\bibitem[\protect\citeauthoryear{{Daigle}, {Carignan} \&
  {Blais-Ouellette}}{{Daigle} et~al.}{2006}]{Dai20066276}
{Daigle} O.,  {Carignan} C.,    {Blais-Ouellette} S.,  2006, in Society of
  Photo-Optical Instrumentation Engineers (SPIE) Conference Series Vol.~6276 of
  Society of Photo-Optical Instrumentation Engineers (SPIE) Conference Series,
  {Faint flux performance of an EMCCD}

\bibitem[\protect\citeauthoryear{{Daigle}, {Carignan}, {Gach}, {Guillaume},
  {Lessard}, {Fortin} \& {Blais-Ouellette}}{{Daigle} et~al.}{2009}]{Dai2009121}
{Daigle} O.,  {Carignan} C.,  {Gach} J.,  {Guillaume} C.,  {Lessard} S.,
  {Fortin} C.,    {Blais-Ouellette} S.,  2009, \pasp, 121, 866

\bibitem[\protect\citeauthoryear{{Daigle}, {Gach}, {Guillaume}, {Carignan},
  {Balard} \& {Boissin}}{{Daigle} et~al.}{2004}]{Dai20045499}
{Daigle} O.,  {Gach} J.,  {Guillaume} C.,  {Carignan} C.,  {Balard} P.,
  {Boissin} O.,  2004, in {J.~D.~Garnett \& J.~W.~Beletic} ed., Society of
  Photo-Optical Instrumentation Engineers (SPIE) Conference Series Vol.~5499 of
  Society of Photo-Optical Instrumentation Engineers (SPIE) Conference Series,
  {L3CCD results in pure photon-counting mode}.
pp 219--227

\bibitem[\protect\citeauthoryear{{Daigle}, {Gach}, {Guillaume}, {Lessard},
  {Carignan} \& {Blais-Ouellette}}{{Daigle} et~al.}{2008}]{Dai20087014}
{Daigle} O.,  {Gach} J.,  {Guillaume} C.,  {Lessard} S.,  {Carignan} C.,
  {Blais-Ouellette} S.,  2008, in Society of Photo-Optical Instrumentation
  Engineers (SPIE) Conference Series Vol.~7014 of Society of Photo-Optical
  Instrumentation Engineers (SPIE) Conference Series, {CCCP: a CCD controller
  for counting photons}

\bibitem[\protect\citeauthoryear{{de Blok} \& {Bosma}}{{de Blok} \&
  {Bosma}}{2002}]{deB2002385}
{de Blok} W.~J.~G.,  {Bosma} A.,  2002, \aap, 385, 816

\bibitem[\protect\citeauthoryear{{de Blok}, {Bosma} \& {McGaugh}}{{de Blok}
  et~al.}{2003}]{deB2003340}
{de Blok} W.~J.~G.,  {Bosma} A.,    {McGaugh} S.,  2003, \mnras, 340, 657

\bibitem[\protect\citeauthoryear{{de Vaucouleurs} \& {Page}}{{de Vaucouleurs}
  \& {Page}}{1962}]{deV1962136}
{de Vaucouleurs} G.,  {Page} J.,  1962, \apj, 136, 107

\bibitem[\protect\citeauthoryear{{Dicaire}, {Carignan}, {Amram}, {Marcelin},
  {Hlavacek-Larrondo}, {de Denus-Baillargeon}, {Daigle} \&
  {Hernandez}}{{Dicaire} et~al.}{2008}]{Dic2008135}
{Dicaire} I.,  {Carignan} C.,  {Amram} P.,  {Marcelin} M.,  {Hlavacek-Larrondo}
  J.,  {de Denus-Baillargeon} M.,  {Daigle} O.,    {Hernandez} O.,  2008, \aj,
  135, 2038

\bibitem[\protect\citeauthoryear{{Dutton}, {Courteau}, {de Jong} \&
  {Carignan}}{{Dutton} et~al.}{2005}]{Dut2005619}
{Dutton} A.~A.,  {Courteau} S.,  {de Jong} R.,    {Carignan} C.,  2005, \apj,
  619, 218

\bibitem[\protect\citeauthoryear{{Epinat}, {Amram}, {Marcelin}, {Balkowski},
  {Daigle}, {Hernandez}, {Chemin}, {Carignan}, {Gach} \& {Balard}}{{Epinat}
  et~al.}{2008}]{Epi2008388}
{Epinat} B.,  {Amram} P.,  {Marcelin} M.,  {Balkowski} C.,  {Daigle} O.,
  {Hernandez} O.,  {Chemin} L.,  {Carignan} C.,  {Gach} J.,    {Balard} P.,
  2008, \mnras, 388, 500

\bibitem[\protect\citeauthoryear{{Ferguson}, {Wyse}, {Gallagher} III \&
  {Hunter}}{{Ferguson} et~al.}{1996}]{Fer1996111}
{Ferguson} A.~M.~N.,  {Wyse} R.~F.~G.,  {Gallagher} III J.~S.,    {Hunter}
  D.~A.,  1996, \aj, 111, 2265

\bibitem[\protect\citeauthoryear{{Gaustad}, {McCullough}, {Rosing} \& {Van
  Buren}}{{Gaustad} et~al.}{2001}]{Gau2001113}
{Gaustad} J.~E.,  {McCullough} P.~R.,  {Rosing} W.,    {Van Buren} D.,  2001,
  \pasp, 113, 1326

\bibitem[\protect\citeauthoryear{{Haffner}, {Dettmar}, {Beckman}, {Wood},
  {Slavin}, {Giammanco}, {Madsen}, {Zurita} \& {Reynolds}}{{Haffner}
  et~al.}{2009}]{Haf200981}
{Haffner} L.~M.,  {Dettmar} R.,  {Beckman} J.~E.,  {Wood} K.,  {Slavin} J.~D.,
  {Giammanco} C.,  {Madsen} G.~J.,  {Zurita} A.,    {Reynolds} R.~J.,  2009,
  Reviews of Modern Physics, 81, 969

\bibitem[\protect\citeauthoryear{{Hlavacek-Larrondo}, {Carignan}, {Daigle}, {de
  Denus-Baillargeon}, {Marcelin}, {Epinat} \& {Hernandez}}{{Hlavacek-Larrondo}
  et~al.}{2010}]{Hla2010}
{Hlavacek-Larrondo} J.,  {Carignan} C.,  {Daigle} O.,  {de Denus-Baillargeon}
  M.,  {Marcelin} M.,  {Epinat} B.,    {Hernandez} O.,  2010, \mnras, pp
  1649--+

\bibitem[\protect\citeauthoryear{{Hoopes}, {Walterbos} \& {Greenwalt}}{{Hoopes}
  et~al.}{1996}]{Hoo1996112}
{Hoopes} C.~G.,  {Walterbos} R.~A.~M.,    {Greenwalt} B.~E.,  1996, \aj, 112,
  1429

\bibitem[\protect\citeauthoryear{{Kassin}, {de Jong} \& {Pogge}}{{Kassin}
  et~al.}{2006}]{Kas2006162}
{Kassin} S.~A.,  {de Jong} R.~S.,    {Pogge} R.~W.,  2006, \apjs, 162, 80

\bibitem[\protect\citeauthoryear{{Kennicutt} Jr., {Lee}, {Funes} Jos{\'e}~G.,
  {Sakai} \& {Akiyama}}{{Kennicutt} et~al.}{2008}]{Ken2008178}
{Kennicutt} Jr. R.~C.,  {Lee} J.~C.,  {Funes} Jos{\'e}~G. S.~J.,  {Sakai} S.,
   {Akiyama} S.,  2008, \apjs, 178, 247

\bibitem[\protect\citeauthoryear{{Marcelin}, {Boulesteix} \&
  {Georgelin}}{{Marcelin} et~al.}{1985}]{Mar1985151}
{Marcelin} M.,  {Boulesteix} J.,    {Georgelin} Y.~P.,  1985, \aap, 151, 144

\bibitem[\protect\citeauthoryear{{Puche} \& {Carignan}}{{Puche} \&
  {Carignan}}{1988}]{Puc198895}
{Puche} D.,  {Carignan} C.,  1988, \aj, 95, 1025

\bibitem[\protect\citeauthoryear{{Puche}, {Carignan} \& {Bosma}}{{Puche}
  et~al.}{1990}]{Puc1990100}
{Puche} D.,  {Carignan} C.,    {Bosma} A.,  1990, \aj, 100, 1468

\bibitem[\protect\citeauthoryear{{Rizzi}, {M{\'e}ndez} \& {Gieren}}{{Rizzi}
  et~al.}{2006}]{Riz2006234}
{Rizzi} L.,  {M{\'e}ndez} R.~H.,    {Gieren} W.,  2006, in {M.~J.~Barlow \&
  R.~H.~M{\'e}ndez} ed., Planetary Nebulae in our Galaxy and Beyond Vol.~234 of
  IAU Symposium, {PNs and H II regions in NGC 300}.
pp 493--494

\bibitem[\protect\citeauthoryear{{Spano}, {Marcelin}, {Amram}, {Carignan},
  {Epinat} \& {Hernandez}}{{Spano} et~al.}{2008}]{Spa2008383}
{Spano} M.,  {Marcelin} M.,  {Amram} P.,  {Carignan} C.,  {Epinat} B.,
  {Hernandez} O.,  2008, \mnras, 383, 297

\bibitem[\protect\citeauthoryear{{van Albada} \& {Sancisi}}{{van Albada} \&
  {Sancisi}}{1986}]{vanA1986320}
{van Albada} T.~S.,  {Sancisi} R.,  1986, Royal Society of London Philosophical
  Transactions Series A, 320, 447

\bibitem[\protect\citeauthoryear{{Vlaji{\'c}}, {Bland-Hawthorn} \&
  {Freeman}}{{Vlaji{\'c}} et~al.}{2009}]{Vla2009697}
{Vlaji{\'c}} M.,  {Bland-Hawthorn} J.,    {Freeman} K.~C.,  2009, \apj, 697,
  361

\end{thebibliography}

\end{document}